\documentclass[11pt,nofootinbib,preprintnumbers,superscriptaddress]{revtex4}

\usepackage{amssymb, amsfonts, amsmath, mathtools, bm}
\usepackage{color}
\usepackage{mathrsfs}
\usepackage{enumerate}
\usepackage{amsfonts}
\usepackage{graphicx}
\usepackage{subcaption}
\usepackage{float}
\usepackage{hyperref}

\begin{document}
\title{
Parametrized Black Hole Quasinormal Ringdown Formalism for Higher Overtones
}

\author{Shin’ichi Hirano}
\affiliation{Oyama National College of Technology, Oyama 323-0806, Japan}
\affiliation{Department of Physics, Rikkyo University, Toshima, Tokyo 171-8501, Japan}

\author{Masashi Kimura}
\affiliation{Department of Information, Artificial Intelligence and Data Science,
Daiichi Institute of Technology, Tokyo 110-0005, Japan}
\affiliation{Department of Physics, Rikkyo University, Toshima, Tokyo 171-8501, Japan}

\author{Masahide Yamaguchi}
\affiliation{Cosmology, Gravity, and Astroparticle Physics Group, Center for Theoretical Physics of the Universe, Institute for Basic Science (IBS), Daejeon, 34126, Korea}
\affiliation{Department of Physics, Tokyo Institute of Technology, Tokyo 152-8551, Japan}

\author{Jiale Zhang}
\affiliation{Department of Physics, Graduate School of Humanities and Sciences,  Ochanomizu University, Tokyo 112-8610, Japan}
\affiliation{Cosmology, Gravity, and Astroparticle Physics Group, Center for Theoretical Physics of the Universe, Institute for Basic Science (IBS), Daejeon, 34126, Korea}

\date{\today}
\preprint{RUP-24-6}

\begin{abstract}

We investigate the parametrized black hole quasinormal ringdown formalism, which is a robust 
framework used to analyze quasinormal modes in systems that closely resemble general relativity, paying particular attention to the higher overtones.
We find that 
larger deviations from the general relativity case 
typically appear in the quasinormal frequencies for the higher overtones.
This growing tendency for higher overtones can be understood using an analytical method, and its relations to previous works are briefly discussed.
Our findings indicate that we can impose a strong constraint on gravity theories by considering the overtones of quasinormal modes.
\end{abstract}

\maketitle
\section{Introduction}

The direct detection of gravitational waves~\cite{LIGOScientific:2016aoc,LIGOScientific:2016vbw}
has led to new perspectives 
in astrophysics, cosmology, and gravity.
In the gravitational wave signal of binary black hole merger, the damped oscillation called the ringdown phase appears after the merger phase, signifying the process of stabilization of the final state black hole.
The spectrum of the ringdown phase is a superposition of damped sinusoids, or quasinormal modes (QNMs), that are uniquely determined by the remnant black hole's mass and spin 
in general relativity (GR).
Through the examination of the ringdown gravitational wave signal, one can estimate the mass and spin of the remnant black hole~\cite{Israel:1967wq,Carter:1971zc,Robinson:1975bv}.

The QNMs in GR are labeled by three indices $(\ell, m, n)$, where $\ell \geq 2$, $-\ell \leq m \leq \ell$, and $n \geq 0$. 
Each set of the $(\ell, m)$ spheroidal-harmonic indices and the overtone $n$ number corresponds to different modes. 
For each set of $(\ell, m)$, $n=0$ modes denote the least-damped mode known as the fundamental mode, and there are an infinite number of overtones \cite{Teukolsky:1972my,Teukolsky:1973ha,Press:1973zz} with $n \geq 1$.
The consistency with the ringdown phase with the 
$n = 0$ fundamental mode has been reported from real data~\cite{LIGOScientific:2016lio, Capano:2020dix, Capano:2021etf}.
In \cite{Giesler:2019uxc}, it was suggested that overtones provide an excellent description of the waveform before the fundamental mode dominates and enable us to extend perturbation theory to times before the peak strain amplitude, yielding surprisingly good fits to numerical relativity simulation. 
This suggests the importance of including overtones in ringdown analysis, and many attempts have been made to fit numerical waveforms while acknowledging the existence of related and crucial arguments~\cite{Correia:2023bfn,Cotesta:2022pci,Mourier:2020mwa,Isi:2020tac,Sago:2021gbq,Oshita:2021iyn,Forteza:2021wfq,MaganaZertuche:2021syq,Cook:2020otn,Dhani:2020nik,Takahashi:2023tkb,Cheung:2023vki,Zhu:2023mzv,Nee:2023osy,Baibhav:2023clw,Mitman:2022qdl,Cheung:2022rbm,Baibhav:2017jhs,London:2014cma,Bhagwat:2019dtm,CalderonBustillo:2020rmh,Dhani:2021vac,Finch:2021iip,Finch:2021qph,Ota:2019bzl,JimenezForteza:2020cve,Ota:2021ypb,Franchini:2023eda}.
Numerous investigations in recent periods have been directed toward isolating QNMs from empirical measurements of gravitational waves~\cite{LIGOScientific:2016lio,LIGOScientific:2019fpa,LIGOScientific:2020tif,LIGOScientific:2021sio,Ghosh:2021mrv}.
Currently, the detection of overtones in real GW data continues to be a subject of debate~\cite{Isi:2019aib,Cotesta:2022pci,Isi:2022mhy,Isi:2023nif,Carullo:2023gtf,Finch:2022ynt,Wang:2023mst,Correia:2023bfn, Wang:2023xsy}.

It is widely believed that inconsistencies of GR will eventually become evident in strong gravity regimes, 
such as near the curvature singularity, or large-scale descriptions \cite{Berti:2015itd,Barack:2018yly}. 
This implies that the gravity theory should be modified from GR at some scale.\footnote{
Although one may assume that the Planck scale is the relevant scale,
no one knows at what scale non-GR effect is important.
Therefore, it is important to verify the theory of gravity from observations without being biased toward new scales.}
The test of gravity theories through ringdown phases can be carried out by comparing the spectra of the ringdown phase with the predictions made in GR or non-GR theories~\cite{Berti:2005ys,Meidam:2014jpa,Berti:2015itd,Berti:2016lat,Berti:2018vdi}.
For this purpose, we utilize the methodology called parametrized black hole quasinormal ringdown formalism (parametrized QNM formalism)~\cite{Cardoso:2019mqo}, where quasinormal frequencies have minor deviations from the predictions of GR in configurations that are perturbatively close to GR. 
Note that while the forms of this formalism assumes the Regge-Wheeler and Zerilli equations, which are master equations for the Schwarzschild case, as the lowest order system, it can treat slowly rotating cases by considering the spin as a perturbative parameter~\cite{Hatsuda:2020egs}.

In this paper, we study the parametrized black hole quasinormal ringdown formalism~\cite{Cardoso:2019mqo} for the higher overtones. 
To calculate the precise values of quasinormal frequencies,
we adopt Nollert's method~\cite{Nollert:1993zz}, which improves convergence
over the larger imaginary parts while dealing with the computational difficulties with overtones.
In contrast to the fundamental mode, we find a growing trend of the model-independent coefficients $e_j$ in the parametrized QNM formalism~\cite{Cardoso:2019mqo} for higher overtones, which is consistent with the $n=1, 2$ overtone cases~\cite{Volkel:2022aca}.
Applying our results to typical modified gravity theories with higher curvature correction terms,
we confirm that the deviation of quasinormal frequencies from the predicted GR value is increased for higher overtones.
We also briefly discuss the case of the algebraically special mode~\cite{Chandrasekhar:1984}, i.e., $n=8$ overtone with $\ell = 2$,
and the relation with recent arguments related to the spectral instability of QNMs~\cite{Jaramillo:2020tuu, Jaramillo:2021tmt, Cheung:2021bol, Berti:2022xfj, Konoplya:2022pbc}.

This paper is organized as follows: 
In section~\ref{sec:pQNMs}, we review the framework of the parametrized QNM formalism along with the calculation methods we adopt. 
In Section~\ref{sec:results}, we show 
our numerical results for the calculation of the coefficients $e_j$ 
in the parametrized QNM formalism 
for higher overtones.
Our analysis reveals a notable trend: the parameter $e_j$ increases with the overtone number $n$. 
This trend is evident in our numerical results and further supported by an analysis of asymptotic behavior of the series $e_j$ discussed in Section~\ref{sec:aymptoticbehavior}. 
In Section~\ref{sec:applications}, we explore the implications of our results for some typical cases.
In Appendix.~\ref{appendixejrecrel},
recursion relations among the coefficients $e_j$ are reviewed.
In Appendix.~\ref{appendix:e1}, we reveal a non-trivial relation including $e_1$. The discussion related to the algebraically special mode $\ell = 2, n=8$ overtone is in Appendix.~\ref{sec:l2n8}. The numerical results of the dimensionless Schwarzschild quasinormal frequencies and the coefficients $e_j$ can be found in Appendices.~\ref{sec:omegadata} and ~\ref{sec:dataej}.

\section{Parametrized Black Hole Quasinormal Ringdown Formalism and calculation method}
\label{sec:pQNMs}

In this section, we briefly review the 
parametrized black hole quasinormal ringdown formalism (parametrized QNM formalism) in~\cite{Cardoso:2019mqo}
and the calculation methods we adopt.
We consider axial (odd-parity) gravitational perturbations whose master equation describes the small deviation
from the Schwarzschild limit in the form
\begin{equation}
f\frac{d}{dr}  \left( f \frac{d\Phi}{dr} \right) + \left(\omega^2 - V\right)\Phi = 0,
\label{eq:mastereqfull}
\end{equation}
where $f = 1- r_H/r$ and $r_H$ denotes the location of the horizon.
The effective potential $V$ is given by
\begin{equation}
   V  = V_{\rm RW} + \delta V,
\label{eq:effectivepotentialfull}
\end{equation}
where $V_{\rm RW}$ is the Regge-Wheeler potential~\cite{Regge:1957td},
\begin{equation}
    V_{\rm RW} = f \left(\frac{\ell(\ell + 1)}{r^2} - \frac{3r_H}{r^3}\right),
    \label{eq:rwpotential}
\end{equation}
 with the notation $\ell$ used as an angular harmonic index to designate the tensorial spherical harmonics,and $\delta V$ describes the small deviations from $V_{\rm RW}$ characterized by the small parameters $\alpha_j$ 
\begin{equation}
    \delta V = \sum_{j=0}^{\infty} \delta V_j = \frac{f}{r_H^2} \sum_{j=0}^{\infty} \alpha_j \left(\frac{r_H}{r}\right)^j.
    \label{eq:effectivepotential}
\end{equation} 
This parametrized potential includes many models~\cite{McManus:2019ulj, Tattersall:2019nmh, Hatsuda:2020egs, deRham:2020ejn, Volkel:2022aca, Volkel:2022khh, Franchini:2022axs, Lahoz:2023csk, Mukohyama:2023xyf}, where different choices of the small parameters $\alpha_j$ correspond to the details of gravitational theories or physical setups (see also Sec.~\ref{sec:applications} for explicit examples).

The quasinormal modes (QNMs) are the solutions of the master equation~\eqref{eq:mastereqfull} that comply with the purely outgoing boundary condition at the spatial infinity~($r_* \to \infty$)\footnote{
If the effective potential has a mass term with the mass squared $\mu^2$, the boundary condition at $r_* \to \infty$ is changed into $\Phi \sim e^{i\sqrt{\omega^2 - \mu^2} r_*}$.
In that case, the ansatz of the wave function used in the series expansion~\eqref{eq:frobeniusseries} 
also should be modified.
}
\begin{equation}
    \Phi \sim e^{i\omega r_*},
\label{eq:BCplus}
\end{equation}
and with the purely ingoing boundary condition
at the horizon~($r_* \to -\infty$)
\begin{equation}
    \Phi \sim e^{-i\omega r_*},
\label{eq:BCminus}
\end{equation} 
where $r_* = r + r_H\ln(r - r_H)/r_H $.
We note that these expressions incorporate the time dependency as $e^{-i\omega t}$.
For the parametrized potentials~\eqref{eq:effectivepotentialfull}-\eqref{eq:effectivepotential}, the corrections in the frequencies are linear in the corrections to the potentials. At the first order of small parameter ${\cal O}(\alpha_j)$, the quasinormal frequencies become~\cite{Cardoso:2019mqo}
\begin{equation}
\omega = \omega_{\rm Sch} + \sum_{j=0}^{\infty} \alpha_j e_j,
    \label{eq:pQNMsomega}
\end{equation}
with
\begin{equation}
    \omega_{\rm Sch}=\frac{2 \Omega_{\rm Sch}}{r_H},\label{eq:omega0}
\end{equation}
where the coefficients $e_j$ are constants which are independent of  $\alpha_j$, and $\Omega_{\rm Sch}$ is the dimensionless Schwarzschild quasinormal frequency normalized by the mass, e.g. $\Omega_{\rm Sch} = 0.3736716844\cdots - i0.0889623157\cdots$ for $\ell=2, n=0$ (see Appendix.~\ref{sec:omegadata} for other $\ell$ and $n$). 
To calculate the quasinormal frequencies in this paper, we employ Leaver's method~\cite{Leaver:1985ax} as well as Nollert's improved method~\cite{Nollert:1993zz}.
The wave function $\Phi$ in Eq.~\eqref{eq:mastereqfull} can be expressed in series form as
\begin{align}
\Phi = e^{i\omega r_*} f^\rho \sum_{k=0}^\infty a_k f^k,
\label{eq:frobeniusseries}
\end{align}
where $\rho = -2ir_H\omega $ is the exponent satisfying the QNM boundary conditions at the horizon $r_* \to -\infty$ in the Frobenius method. 
The master equation can be reduced to recursion relations for the expansion coefficients $a_k$ after performing some calculations.
If a solution such that the summation in \eqref{eq:frobeniusseries} at infinity is convergent, i.e., 
$\sum_{k=0}^\infty a_k$ is convergent, can be found, the QNM boundary condition at the spatial infinity $r_* \to \infty$ is simultaneously satisfied.
Such frequencies that lead to convergent summation $\sum_{k=0}^\infty a_k$ are the quasinormal frequencies.
We can find these frequencies using so-called Leaver's continuous fraction method with high accuracy, which is mathematically equivalent to solving the equation $a_{K} = 0$ for a large integer $K$.
In fact, to obtain the quasinormal frequencies for higher overtones with high accuracy,
we need Nollert's improved method, where we solve
\begin{align}
\frac{a_K}{a_{K-1}} =\mathbf{c}_0 +\mathbf{c}_1 K^{-1/2}+\mathbf{c}_2 K^{-1}+\cdots,
\label{eq:nollertcj}
\end{align}
instead of $a_{K} = 0$.
In Eq.~\eqref{eq:nollertcj}, 
the coefficients $\mathbf{c}_i$ are determined from the asymptotic form of the recursion relation for $a_k$ at large $k$ 
with the convergence condition of the summation of $a_k$~\cite{Nollert:1993zz}.
In Leaver's method, we only impose that the summation of $a_k$ is convergent, 
whereas, in Nollert's method, we also take into account the convergent rate of $a_k$.
For this reason, solving Eq.~\eqref{eq:nollertcj} can 
give more accurate quasinormal frequencies than 
solving $a_{K} = 0$ for a fixed integer $K$.
The technical details of Leaver's method and Nollert's method can be found in~\cite{Konoplya:2011qq}.

To determine the numerical value of $e_j$ in Eq.~\eqref{eq:pQNMsomega}, 
we use the perturbative calculation technique developed in~\cite{Hatsuda:2023geo},
i.e., we solve Eq.~\eqref{eq:nollertcj} at the order ${\cal O}(\alpha_j)$ after substituting Eq.~\eqref{eq:pQNMsomega} into Eq.~\eqref{eq:nollertcj}. The result is summarized in the next section.

\section{Results of numerical calculation for $e_j$}
\label{sec:results}
We calculated the precise numerical values of the coefficients of the parametrized QNM formalism
$e_j$ in Eq.~\eqref{eq:pQNMsomega} for higher overtones using Leaver's method~\cite{Leaver:1985ax} and Nollert's improved method~\cite{Nollert:1993zz}.
The results of numerical calculations are detailed in Appendix.~\ref{sec:dataej}, with the case of $n=8$ for $\ell=2$ being a special scenario discussed in Appendix.~\ref{sec:l2n8}.
In Appendix.~\ref{sec:omegadata}, we also present the numerical values of quasinormal frequencies for Schwarzschild black holes.
Our numerical data are available on the website~\cite{url:pqnm_overtone}.

We checked the accuracy of our numerically calculated coefficients $e_j$ by
substituting them into Eq.~\eqref{eq:ejrecrelation}, the identical relations for $e_j$ \cite{Kimura:2020mrh}.
Defining the function of error estimation as
\begin{align}
\Delta_{e_j} = 
  \frac{|c_{j+1} e_{j+1}+ c_{j+3} e_{j+3} + c_{j+4} e_{j+4} + c_{j+5}e_{j+5}|}{
|c_{j+1} e_{j+1}|+ |c_{j+3} e_{j+3}| + |c_{j+4} e_{j+4}| + |c_{j+5}e_{j+5}|
},
   \label{eq:ejerror}
\end{align}
where $c_{j+a}$ are given by Eqs.~\eqref{eq:cj1}-\eqref{eq:cj5},
we confirmed that 
the margin of error
Eq.~\eqref{eq:ejerror} 
is at most 
${\cal O}(10^{-21})$ for our numerical data of $e_j$.

Table~\ref{tab:selected_ej_values} shows the typical values of $e_j$ for the fundamental mode and overtones. 
While the absolute value of $e_j$ for the fundamental mode ($n=0$) decays as $j$ increases, it grows for overtones as $j$ increases in the large $j$ regime.
This suggests that higher overtone modes are much more sensitive to modifications in the effective potential, which is consistent with \cite{Volkel:2022aca}. 
The growing trend of the coefficients $e_j$ for large $j$ in overtones can be understood from the asymptotic analysis 
as discussed in the next section.

\begin{table}[t]
\centering
\caption{Typical numerical values of $e_j$. 
Cases for $\ell =2$ and $n=0, 2, 10$ are presented for various values of $j$.
}
\label{tab:selected_ej_values}
\vspace{-10pt}
\begin{subtable}{0.33\textwidth}
\caption{$n=0$}
\label{tab:selected_ej_values_n0}
\begin{tabular}{cc}
\hline
$j$ & $r_H e_j$ \\
\hline
0 & $0.24725 + 0.092643i$ \vspace{-4pt}\\
1 & $0.15985 + 0.018208i$ \vspace{-4pt}\\
2 & $0.096632 - 0.0024155i$ \vspace{-4pt}\\
10 & $0.0036853 + 0.0065244i$ \vspace{-4pt}\\
15 & $-0.00081238 + 0.0047307i$ \vspace{-4pt}\\
20 & $-0.0022183 + 0.0029672i$ \vspace{-4pt}\\
25 & $-0.0025952 + 0.0016658i$ \vspace{-4pt}\\
50 & $-0.0016013 - 0.00090149i$ \vspace{-4pt}\\
75 & $-0.00058591 - 0.0012547i$ \vspace{-4pt}\\
100 & $6.4338\times10^{-7} - 0.0011385i$ \\
\hline
\end{tabular}
\end{subtable}
\hfill 
\hspace*{-10pt}
\begin{subtable}{0.33\textwidth}
\caption{$n=2$}
\label{tab:selected_ej_values_n2}
\begin{tabular}{cc}
\hline
$j$ & $r_H e_j$ \\
\hline
0 & $-0.006678 + 0.14146i$ \vspace{-4pt}\\
1 & $0.11931 + 0.075616i$ \vspace{-4pt}\\
2 & $0.11467 + 0.00074885i$ \vspace{-4pt}\\
10 & $0.048075 + 0.052281i$ \vspace{-4pt}\\
15 & $0.029569 + 0.081252i$ \vspace{-4pt}\\
20 & $0.0057647 + 0.10267i$ \vspace{-4pt}\\
25 & $-0.021305 + 0.11745i$ \vspace{-4pt}\\
50 & $-0.16315 + 0.11792i$  \vspace{-4pt}\\
75 & $-0.27727 + 0.045252i$ \vspace{-4pt}\\
100& $-0.35334 - 0.061499i$  \\
\hline
\end{tabular}
\end{subtable}
\hfill
\hspace*{-4pt}
\begin{subtable}{0.33\textwidth}
\caption{$n=10$}
\label{tab:selected_ej_values_n10}
\begin{tabular}{cc}
\hline
$j$ & $r_H e_j$ \\
\hline
0 & $-0.028643 + 0.010611i$ \vspace{-4pt}\\
1 & $-0.044125 + 0.046418i$ \vspace{-4pt}\\
2 & $-0.075205 + 0.046894i$ \vspace{-4pt}\\
10 & $-11.971 + 1.9125i$ \vspace{-4pt}\\
15 & $-82.298 + 9.0573i$ \vspace{-4pt}\\
20 & $-390.83 + 28.087i$ \vspace{-4pt}\\
25 & $-1459.5 + 57.673i$ \vspace{-4pt}\\
50 & $-1.5199\times10^{5} - 12603i$ \vspace{-5pt} \\
75 & $-3.1869\times10^{6} - 5.4383\times10^{5}i$ \vspace{-4pt}\\
100& $-3.0889\times10^{7} - 7.4294\times10^{6}i$\\
\hline
\end{tabular}
\end{subtable}

\end{table}

\section{Asymptotic behavior of $e_j$ for large $j$}
\label{sec:aymptoticbehavior}

In this section, we discuss the asymptotic behavior of the coefficients $e_j$ for large $j$ .
As discussed in Appendix.~\ref{appendixejrecrel}, the coefficients $e_j$ satisfy the recursion relations in Eq.~\eqref{eq:ejrecrelation}.
By defining the ratio of $e_j$ and $e_{j-1}$ 
as $b_j := e_j/e_{j-1}$, Eq.~\eqref{eq:ejrecrelation} has an approximate solution for large $j$ in the Taylor series as 
\begin{equation}
    b_j  = \frac{e_j}{e_{j-1}} 
    = 1 + \sum_{i = 1}^\infty \frac{C_i}{j^i},
    \label{eq:asymptoticbj}
\end{equation}
where 
$C_i$ are constants that capture the behavior of the series.
The explicit forms of $C_i$ with lower indices $i$ are\footnote{
Since Eq.~\eqref{eq:ejrecrelation} represents a five-term recursion relation, the general solution is described by a superposition of four fundamental solutions. The fundamental solution which becomes the leading term in the general solution for large $j$ satisfies the ansatz Eq.~\eqref{eq:asymptoticbj} with Eq.~\eqref{eq:constc1}.
We note that the contribution of the other three solutions is subdominant 
in the large $j$ regime compared to the solution Eq.~\eqref{eq:asymptoticbj} with Eq.~\eqref{eq:constc1}.
}

\begin{align}
C_1 &= i (i + 2 r_H \omega_{\rm Sch}), 
\label{eq:constc1}
\\
C_2 &= -C_1^2 - 2 (-3 + \ell + \ell^2), \\
C_3 &= 2 C_1^3 - 6 (-3 + \ell + \ell^2) + \frac{2 (-3 + \ell + \ell^2)^2}{C_1} + C_1 (-1 + 2 \ell (1 + \ell)), \\
C_4 &= C_1^3 - 5 C_1^4 - 38 (-3 + \ell + \ell^2) + \frac{18 (-3 + \ell + \ell^2)^2}{C_1} \nonumber \\
&\quad- 2 C_1^2 (-5 + 3 \ell + 3 \ell^2) + C_1 (-1 + 6 \ell + 6 \ell^2), \\
C_5 &= \frac{1}{(-1 + C_1) C_1^2} (-21 C_1^7 + 14 C_1^8 - 2 (-3 + \ell + \ell^2)^4 - 14 C_1^5 (-5 + 3 \ell + 3 \ell^2) \nonumber \\
&\quad + 4 C_1^6 (-8 + 5 \ell + 5 \ell^2) - 2 C_1 (-3 + \ell + \ell^2)^2 (55 + 6 \ell + 6 \ell^2) \nonumber \\
&\quad - 2 C_1^3 (-284 + 103 \ell + 106 \ell^2 + 6 \ell^3 + 3 \ell^4) + C_1^4 (-23 + 30 \ell + 36 \ell^2 + 12 \ell^3 + 6 \ell^4) \nonumber \\
&\quad - 2 C_1^2 (-252 + 309 \ell + 228 \ell^2 - 160 \ell^3 - 75 \ell^4 + 6 \ell^5 + 2 \ell^6)), 
\nonumber \\
C_6 &= -\frac{1}{(-1+C_1) C_1^2} \left(-79 C_1^8 + 42 C_1^9 + 32 (-3+\ell+\ell^2)^4 + 35 C_1^7 (-3+2 \ell+2 \ell^2) \right. \nonumber \\
&\quad - 4 C_1^6 (-77+41 \ell+41 \ell^2) + 2 C_1 (-3+\ell+\ell^2)^2 (277+96 \ell+96 \ell^2) \nonumber \\
&\quad + C_1^5 (-173+64 \ell+96 \ell^2+64 \ell^3+32 \ell^4) 
- 2 C_1^4 (10-22 \ell+17 \ell^2+78 \ell^3+39 \ell^4) \nonumber \\
&\quad + C_1^3 (-2643+960 \ell+1010 \ell^2+100 \ell^3+50 \ell^4) \nonumber \\
&\quad \left. + 2 C_1^2 (-1878+2423 \ell+1728 \ell^2-1358 \ell^3-599 \ell^4+96 \ell^5+32 \ell^6)\right), \\
C_7 &= \frac{1}{(-2+C_1) (-1+C_1) C_1^3}\left(-572 C_1^{11} + 132 C_1^{12} + 4 (-3+\ell+\ell^2)^6 \right. \nonumber \\
&\quad + 8 C_1 (-3+\ell+\ell^2)^4 (76+5 \ell+5 \ell^2)  
+ 4 C_1^{10} (74+63 \ell+63 \ell^2) - 4 C_1^9 (-493+295 \ell+295 \ell^2) \nonumber \\
&\quad + 2 C_1^8 (-1752+778 \ell+853 \ell^2+150 \ell^3+75 \ell^4) 
- 2 C_1^7 (-695-200 \ell+162 \ell^2+724 \ell^3+362 \ell^4) \nonumber \\
&\quad + 4 C_1^2 (-3+\ell+\ell^2)^2 (487+1491 \ell+1414 \ell^2-151 \ell^3-68 \ell^4+9 \ell^5+3 \ell^6) \nonumber \\
&\quad + C_1^6 (967-2346 \ell-1034 \ell^2+2644 \ell^3+1372 \ell^4+60 \ell^5+20 \ell^6) \nonumber \\
&\quad - C_1^5 (-10725+2772 \ell+4016 \ell^2+2540 \ell^3+1400 \ell^4+156 \ell^5+52 \ell^6) \nonumber \\
&\quad - 2 C_1^4 (-1416+12583 \ell+7095 \ell^2-10678 \ell^3-4594 \ell^4+894 \ell^5+298 \ell^6) \nonumber \\
&\quad \left. + 2 C_1^3 (-34911+32394 \ell+27233 \ell^2-11040 \ell^3-7271 \ell^4-1978 \ell^5-454 \ell^6+176 \ell^7+44 \ell^8)\right).
\end{align}
Eq.~\eqref{eq:asymptoticbj} can be solved with respect to $e_j$ as
\begin{equation}
    e_j \simeq C e^{A_1 \ln j - 1  + P(j)},
    \label{eq:ejseries}
\end{equation}
where $C$ is a constant and the function $P(j)$ is given by
\begin{equation}
 P(j) = 1 + \sum_{i=1}^{\infty}\frac{A_{i+1}}{j^i},
 \label{eq:PJ}
\end{equation}
wherein the coefficients $A_i$ can be defined through their correlations with $C_i$:
\begin{align}
    A_1 &= C_1, \\
    A_2 &= \frac{1}{2}(C_1 + C_1^2 - 2C_2), \\
    A_3 &= -\frac{1}{12}(1 + C_1)(C_1 + 2C_1^2 - 6C_2) - \frac{C_3}{2}, \\
    A_4 &= \frac{1}{12}(2C_1^3 + C_1^4 + C_1^2(1 - 4C_2) + 2(-1 + C_2)C_2 
    + 6C_3 + C_1(-6C_2 + 4C_3) - 4C_4), \\
    A_5 &= \frac{1}{120}(-15C_1^4 - 6C_1^5 + 10C_1^3(-1 + 3C_2) + 30C_1^2(2C_2 - C_3) \nonumber\\
    &\quad + C_1(1 - 30(-1 + C_2)C_2 - 60C_3 + 30C_4) - 30(C_2^2 + C_3 - C_2C_3 - 2C_4 + C_5)),\\
    A_6 &= \frac{1}{60} \left( 6 C_1^5 + 2 C_1^6 + C_1^4 (5 - 12 C_2) + 6 C_1^3 (-5 C_2 + 2 C_3) \right. \nonumber \\
    &\quad \left. + C_1^2 (-1 + 2 C_2 (-10 + 9 C_2) + 30 C_3 - 12 C_4) \right. 
    \left. + 2 C_1 (3 C_2 (5 C_2 - 4 C_3) + 10 C_3 - 15 C_4 + 6 C_5) \right. \nonumber \\
    &\quad \left. + 2 (C_2 + 5 C_2^2 - 2 C_2^3 - 15 C_2 C_3 + 3 C_3^2 - 10 C_4 + 6 C_2 C_4 + 15 C_5 - 6 C_6) \right), \\
    A_7 &= \frac{1}{252} \left( -21 C_1^6 - 6 C_1^7 + 21 C_1^5 (-1 + 2 C_2) + 42 C_1^4 (3 C_2 - C_3) \right. \nonumber \\
    &\quad \left. + 7 C_1^3 (1 + 3 (5 - 4 C_2) C_2 - 18 C_3 + 6 C_4) - 21 C_1^2 (9 C_2^2 + 5 C_3 - 6 C_2 C_3 - 6 C_4 + 2 C_5) \right. \nonumber \\
    &\quad \left. + C_1 (-1 + 21 (-5 C_2^2 + 2 C_2^3 - 2 C_3^2 + C_2 (-1 + 12 C_3 - 4 C_4)) + 105 C_4 - 126 C_5 + 42 C_6) \right. \nonumber \\
    &\quad \left. + 21 (2 C_2^3 + C_3 - 2 C_2^2 C_3 - 3 C_3^2 + 2 C_3 C_4 - 5 C_5 + C_2 (5 C_3 - 6 C_4 + 2 C_5) + 6 C_6 - 2 C_7) \right).
\end{align}
Hence, the leading term of $e_j$ becomes\footnote{Eq.~\eqref{eq:leadingterm} is valid for any value of the overtone number $n$ and the angular harmonic index $\ell$, as long as $j$ is sufficiently large.} 
\begin{equation}
    e_j \simeq e_j^{\rm Leading} =  C e^{A_1 \ln j} = C j^{-1 - 2r_H \omega_I} e^{2ir_H \omega_R \ln j},
    \label{eq:leadingterm}
\end{equation}
where $\omega_{\rm Sch} = \omega_R + i\omega_I$.
This coincides with the asymptotic form of $e_j$ estimated in \cite{Cardoso:2019mqo}.
This demonstrates that the factor $j^{-1 - 2r_H \omega_I}$ determines the convergence or divergence of $|e_j|$ for large $j$.
In the left three panels of Table~\ref{tab:powerofej}, we display the values of $j^{-1 - 2r_H \omega_I}$ for $\ell = 2,3,4$ across various overtone numbers $n$. 
For the fundamental mode, $|e_j|$ shows a decreasing trend at $j \to \infty$; yet, it shows divergence for overtones. 
This indicates that the correction to the QNM frequency $\alpha_j e_j$, as outlined in Eq.~\eqref{eq:pQNMsomega} contributes significantly to overtones at large $j$.

Let us consider the physical implications of our results.
For large $j$, 
$\delta V_j$ in Eq.~\eqref{eq:effectivepotential} 
has a peak near the horizon
\begin{align}
r_* = - r_H \ln(j/e),
\end{align}
since $\delta V_j$ for large $j$ near the peak can be written in the form
\begin{align}
\delta V_j = f \frac{\alpha_j}{r_H^2} \left(\frac{r_H}{r}\right)^j \simeq \frac{1}{e r_H^2}\frac{\alpha_j}{j}e^{-(r_* + r_H \ln(j/e))^2/(2r_H^2)},
\label{eq:deltavjlargej}
\end{align}
as demonstrated in~\cite{Cardoso:2019mqo}.
This indicates that the large $j$ limit corresponds to modifications to the effective potential in the near-horizon regime. On the other hand, we confirmed the increasing trend of $e_j$ for overtones as $j$ increases, in contrast to the fundamental mode where $e_j$ decays as $j$ increases.\footnote{
If we define $\tilde{\alpha}_j := \alpha_j / j$ and $\tilde{e}_j := j e_j$ for $j \ge 1$,
the amplitude of $\delta V_j$ in Eq.~\eqref{eq:deltavjlargej} 
is proportional to $\tilde{\alpha}_j$, 
and the modification to the quasinormal frequency
in Eq.~\eqref{eq:pQNMsomega}
becomes $\sum_j \alpha_j e_j = \sum_j \tilde{\alpha}_j \tilde{e}_j$.
From Eq.~\eqref{eq:leadingterm}, 
$\tilde{e}_j$ for large $j$ behaves
$\tilde{e}_j \simeq  C j^{- 2r_H \omega_I} e^{2ir_H \omega_R \ln j}$.
While the factor $j^{- 2r_H \omega_I}$ grows as $j$ increases for both the fundamental mode and overtones,
the factor for the fundamental mode is still relatively smaller than that for the overtones.
}
Thus, we conclude that near-horizon modifications of the effective potential strongly influence the overtones of QNMs but not the fundamental mode.
This is 
consistent with results in~\cite{Konoplya:2022pbc} which states that the first few overtones 
are capable of probing the geometry 
close to the event horizon.
Also, our result provides a partial answer to the recently confirmed findings
that the overtone spectra are relatively more unstable against perturbations in the effective potential compared to the fundamental mode~\cite{Jaramillo:2020tuu, Jaramillo:2021tmt, Cheung:2021bol, Berti:2022xfj}.

\begin{table}[t]
\caption{
Values 
of $j^{-1 - 2r_H \omega_I}$ for $\ell = 2,3,4$ 
and the WKB formula 
across various overtone numbers $n$. 
The WKB formula in Eq.~\eqref{eq:wkbomega} corresponds to the approximation for large $\ell$.
}
\vspace{-10pt}
\begin{center}

\begin{subtable}{.24\linewidth}
\centering
\caption{$\ell = 2$}
\begin{tabular}{cc}
\hline
$n$ &  $j^{-1 - 2r_H \omega_I}$\\
\hline
$0$ &  $j^{-0.64415074}$ \vspace{-4pt}\\
$1$ &  $j^{0.095659501 }$ \vspace{-4pt}\\
$2$ &  $j^{0.91310793 }$ \vspace{-4pt}\\
$3$ &  $j^{1.8205928 }$ \vspace{-4pt}\\
$4$ &  $j^{2.7873796 }$ \vspace{-4pt}\\
$5$ &  $j^{3.7824322 }$ \vspace{-4pt}\\
$6$ &  $j^{4.7916425 }$ \vspace{-4pt}\\
$7$ &  $j^{5.8153647 }$\\
\hline
\end{tabular}
\end{subtable}
\hfill
\begin{subtable}{.24\linewidth}
\centering
\caption{$\ell = 3$}
\begin{tabular}{cc}
\hline
$n$ &  $j^{-1 - 2r_H \omega_I}$\\
\hline
$0$ &  $j^{-0.62918781}$ \vspace{-4pt}\\
$1$ &  $j^{0.12519245 }$ \vspace{-4pt}\\
$2$ &  $j^{0.91637100 }$ \vspace{-4pt}\\
$3$ &  $j^{1.7613484 }$ \vspace{-4pt}\\
$4$ &  $j^{2.6625976 }$ \vspace{-4pt}\\
$5$ &  $j^{3.6086054 }$ \vspace{-4pt}\\
$6$ &  $j^{4.5836490 }$ \vspace{-4pt}\\
$7$ &  $j^{5.5753781 }$ \\
\hline
\end{tabular}
\end{subtable}
\hfill
\begin{subtable}{.24\linewidth}
\centering
\caption{$\ell = 4$}
\begin{tabular}{cc}
\hline
$n$ &  $j^{-1 - 2r_H \omega_I}$\\
\hline
$0$ &  $j^{-0.62334416}$ \vspace{-4pt}\\
$1$ &  $j^{0.13733740 }$ \vspace{-4pt}\\
$2$ &  $j^{0.91963270 }$ \vspace{-4pt}\\
$3$ &  $j^{1.7356973 }$ \vspace{-4pt}\\
$4$ &  $j^{2.5929559 }$ \vspace{-4pt}\\
$5$ &  $j^{3.4919070 }$ \vspace{-4pt}\\
$6$ &  $j^{4.4267451 }$ \vspace{-4pt}\\
$7$ &  $j^{5.3886827 }$ \\
\hline
\end{tabular}
\end{subtable}
\hfill
\begin{subtable}{.24\linewidth}
\centering
\caption{$\text{WKB}$ (large $\ell$)}
\begin{tabular}{cc}
\hline
$n$ &  $j^{-1 - 2r_H \omega_I}$\\
\hline
0 & $ j^{-0.61509982 } $ \vspace{-4pt}\\
1 & $ j^{0.15470054 } $ \vspace{-4pt}\\
2 & $ j^{0.92450090 } $ \vspace{-4pt}\\
3 & $ j^{1.6943013 } $ \vspace{-4pt}\\
4 & $ j^{2.4641016 } $ \vspace{-4pt}\\
5 & $ j^{3.2339020 } $ \vspace{-4pt}\\
6 & $ j^{4.0037023 } $ \vspace{-4pt}\\
7 & $ j^{4.7735027 } $ \\
\hline
\end{tabular}
\end{subtable}
\end{center}
\label{tab:powerofej}
\end{table}%

For the large $\ell$ case, it is known that the first-order Wentzel-Kramers-Brillouin (WKB) formula
in~\cite{Schutz:1985km}
\begin{align}
\omega_{\rm Sch} \simeq 
\omega_{\rm WKB} = \frac{2}{3\sqrt{3}r_H}\left(
\ell + \frac{1}{2}  - i \left(n+\frac{1}{2}\right) \right)
+{\cal O}(1/\ell),
\label{eq:wkbomega}
\end{align}
provides a good approximation of the quasinormal frequency for the Schwarzschild black hole.
In the right panel of Table~\ref{tab:powerofej}, we also
show the value of $j^{-1 - 2r_H \omega_I}$ for the WKB formula using Eq.~\eqref{eq:wkbomega}.
We observe that the trend is consistent with the cases involving lower $\ell$ values.

To extend the validity of the approximate solution~\eqref{eq:ejseries} with Eq.~\eqref{eq:PJ}
for large $j$ to lower values of $j$, we applied the Pad\'e approximants to the 
function $P(j)$ 
up to $j^{-6}$ order, which can be written as follows:
\begin{equation}
P^{[3/3]} = 
\frac{\tilde{A}_0 + \tilde{A}_1 /j + \tilde{A}_2 /j^2 + \tilde{A}_3 /j^3}{1 + \tilde{B}_1/j + \tilde{B}_2 /j^2 +\tilde{B}_3 /j^3},
\label{eq:pade33}
\end{equation}
where the coefficients $\tilde{A}_i, \tilde{B}_i$ are selected so that the Taylor series of Eq.~\eqref{eq:pade33} matches that of $P(j)$ in Eq.~\eqref{eq:PJ} up to $j^{-6}$.
Replacing $P(j)$ in Eq.~\eqref{eq:ejseries} with $P^{[3/3]}$, we obtain the approximate expression
\begin{align}
e_j^{\text{Pad\'e}} = C e^{A_1 \ln j - 1  + P^{[3/3]}}.
\label{eq:ejpadeapp}
\end{align}
Table~\ref{tab:ejRatios}
shows values of $e_{j}/e_{j-1}$ for our numerical results, 
$e_j^{\rm Leading}$ in Eq.\eqref{eq:leadingterm} 
and $e_j^{\text{Pad\'e}}$ in Eq.~\eqref{eq:ejpadeapp} in cases $\ell = 2, n = 2, 10$.

Both the approximate expressions derived from the leading behavior and the Pad\'e approximants are consistent with the numerical results for large values of $j$. However, only the Pad\'e approximants exhibit a high level of agreement with the numerical results even for small values of $j$.
We confirmed that the same tendency can be observed across various values of angular harmonic index $\ell$ and overtone numbers $n$.

In the above discussion, we show that $|e_j|$ with high $j$ increases for higher overtones.
This raises the question of whether the quasinormal frequencies are still described by small corrections to the GR case.
Therefore, we comment on the validity of the parametrized QNM formalism in Eq.~\eqref{eq:pQNMsomega}.
For the parametrized potential in Eq.~\eqref{eq:effectivepotential}, the quasinormal frequency $\omega$
is described by the function of the parameters $\alpha_j$.
Assuming that the quasinormal frequency $\omega$ can be expanded in a Taylor series of $\alpha_j$, 
we obtain Eq.~\eqref{eq:pQNMsomega} by taking up to the the first order. 
The first order approximation~\eqref{eq:pQNMsomega} 
is a good approximation if 
\begin{align}
|\omega_{\rm Sch}| \gg \sum_{j=0}^{\infty} |\alpha_j e_j|,
\label{eq:validityofexpansion}
\end{align}
is satisfied.
Thus, as far as we consider sufficiently small parameters $\alpha_j$, the expansion in Eq.~\eqref{eq:pQNMsomega}
is still a good approximation even if $|e_j|$ takes very large value.

On the other hand, if we consider fixed values of the parameters $\alpha_j$, even though the expansion in Eq.~\eqref{eq:pQNMsomega} is valid for lower overtones, we should not expect it to be a good approximation for higher overtones.
The threshold of the validity of the expansion in~Eq.~\eqref{eq:pQNMsomega} can be estimated by finding overtone number $n$ which gives
\begin{align}
|\omega_{\rm Sch}| \sim \sum_{j=0}^{\infty} |\alpha_j e_j|.
\label{eq:thresholdestimation}
\end{align}
In the next section, we discuss the threshold for some examples.

\begin{table}[t]
\centering
\caption{Values of $e_{j}/e_{j-1}$ for $\ell = 2$ with different overtone numbers $n$.}
\label{tab:ejRatios}
\hspace*{-20pt}
\begin{subtable}{0.49\textwidth}
\centering
\small
\caption{$n = 2$}
\label{tab:ejRatio_n2}
\begin{tabular}{cccc}
\hline
$j$ & Numerics & Leading & Pad\'e  \\
\hline
3 & $0.6820 - 0.1190i$ & $1.279 + 0.679i$ & $0.2616 + 0.4728i$ \vspace{-4pt}\\
4 & $0.7421 + 0.1408i$ & $1.2232 + 0.4415i$ & $0.7046 + 0.2915i$ \vspace{-4pt}\\
5 & $0.9493 + 0.2468i$ & $1.1820 + 0.3255i$ & $0.8791 + 0.2180i$ \vspace{-4pt}\\
6 & $1.0484 + 0.1552i$ & $1.1528 + 0.2572i$ & $0.9586 + 0.1835i$ \vspace{-4pt}\\
7 & $1.0341 + 0.1073i$ & $1.1314 + 0.2125i$ & $1.0012 + 0.1603i$ \vspace{-4pt}\\
8 & $1.0221 + 0.1034i$ & $1.1151 + 0.1809i$ & $1.0253 + 0.1392i$ \vspace{-4pt}\\
9 & $1.0244 + 0.1050i$ & $1.1024 + 0.1574i$ & $1.0368 + 0.1200i$ \vspace{-4pt}\\
10 & $1.0309 + 0.1016i$ & $1.0921 + 0.1393i$ & $1.0404 + 0.1049i$ \vspace{-4pt}\\
11 & $1.0355 + 0.0948i$ & $1.0838 + 0.1249i$ & $1.0405 + 0.0941i$ \vspace{-4pt}\\
12 & $1.0375 + 0.0875i$ & $1.0768 + 0.1132i$ & $1.0396 + 0.0862i$ \vspace{-4pt}\\
13 & $1.0377 + 0.0810i$ & $1.0708 + 0.1035i$ & $1.0385 + 0.0800i$ \vspace{-4pt}\\
14 & $1.0372 + 0.0756i$ & $1.0658 + 0.0954i$ & $1.0374 + 0.0750i$ \vspace{-4pt}\\
15 & $1.0363 + 0.0710i$ & $1.0614 + 0.0884i$ & $1.0364 + 0.0706i$ \vspace{-4pt}\\
16 & $1.0354 + 0.0670i$ & $1.0575 + 0.0824i$ & $1.0354 + 0.0668i$ \vspace{-4pt}\\
17 & $1.0344 + 0.0635i$ & $1.0541 + 0.0771i$ & $1.0344 + 0.0634i$ \vspace{-4pt}\\
18 & $1.0335 + 0.0603i$ & $1.0511 + 0.0725i$ & $1.0335 + 0.0603i$ \vspace{-4pt}\\
19 & $1.0325 + 0.0575i$ & $1.0484 + 0.0684i$ & $1.0325 + 0.0574i$ \vspace{-4pt}\\
20 & $1.0316 + 0.0549i$ & $1.0460 + 0.0647i$ & $1.0316 + 0.0549i$ \\
\hline
\end{tabular}
\end{subtable}
\hfill
\begin{subtable}{0.49\textwidth}
\centering
\caption{$n = 10$}
\label{tab:ejRatio_n10}
\begin{tabular}{cccc}
\hline
$j$ & Numerics  & Leading  & Pad\'e \\
\hline
3 & $2.431 + 0.955i$ & $42.10 + 5.25i$ & $4.048 + 0.268i$ \vspace{-4pt}\\
4 & $2.251 - 0.103i$ & $14.23 + 1.26i$ & $2.570 + 0.096i$ \vspace{-4pt}\\
5 & $1.807 - 0.020i$ & $7.848 + 0.537i$ & $2.111 + 0.053i$ \vspace{-4pt}\\
6 & $1.772 + 0.107i$ & $5.385 + 0.301i$ & $1.895 + 0.035i$ \vspace{-4pt}\\
7 & $1.781 + 0.040i$ & $4.153 + 0.196i$ & $1.768 + 0.027i$ \vspace{-4pt}\\
8 & $1.674 - 0.024i$ & $3.433 + 0.140i$ & $1.683 + 0.022i$ \vspace{-4pt}\\
9 & $1.578 - 0.006i$ & $2.968 + 0.107i$ & $1.619 + 0.019i$ \vspace{-4pt}\\
10 & $1.539 + 0.026i$ & $2.647 + 0.085i$ & $1.570 + 0.017i$ \vspace{-4pt}\\
11 & $1.525 + 0.029i$ & $2.412 + 0.070i$ & $1.530 + 0.015i$ \vspace{-4pt} \\
12 & $1.501 + 0.016i$ & $2.234 + 0.060i$ & $1.496 + 0.014i$ \vspace{-4pt}\\
13 & $1.467 + 0.008i$ & $2.095 + 0.051i$ & $1.466 + 0.013i$ \vspace{-5pt} \\
14 & $1.437 + 0.008i$ & $1.983 + 0.045i$ & $1.441 + 0.012i$ \vspace{-4pt}\\
15 & $1.415 + 0.011i$ & $1.892 + 0.040i$ & $1.418 + 0.012i$ \vspace{-4pt}\\
16 & $1.3964 + 0.0119i$ & $1.815 + 0.036i$ & $1.3984 + 0.0110i$ \vspace{-4pt}\\
17 & $1.3797 + 0.0112i$ & $1.751 + 0.033i$ & $1.3803 + 0.0105i$ \vspace{-4pt}\\
18 & $1.3638 + 0.0101i$ & $1.696 + 0.030i$ & $1.3640 + 0.0101i$ \vspace{-4pt}\\
19 & $1.3488 + 0.0095i$ & $1.648 + 0.027i$ & $1.3491 + 0.0097i$ \vspace{-4pt}\\
20 & $1.3351 + 0.0091i$ & $1.606 + 0.025i$ & $1.3354 + 0.0093i$ \\
\hline
\end{tabular}
\end{subtable}
\end{table}

\section{Applications}
\label{sec:applications}
In this section, we apply our formalism to typical gravity theories with higher curvature correction terms and consider the slowly rotating Kerr black hole cases.
\subsection{Effective field theory extension of GR
}\label{sec:EFT}

The analysis of black hole models developed within the framework of effective field theory (EFT), as discussed in the studies~\cite{Endlich:2017tqa,Cardoso:2018ptl,Cano:2019ore,deRham:2020ejn,Nomura:2021efi,Cano:2022wwo,Silva:2022srr,Cano:2023jbk,Cayuso:2023xbc,Cano:2023tmv,Melville:2024zjq}. 
Refs. \cite{Endlich:2017tqa} and \cite{Cardoso:2018ptl} are representative examples in this context to investigate our formalism, where the model differs from GR through three coupling constants that determine the extent of modifications at high curvature levels.  
For simplicity, our discussion will be limited to a single coupling constant, represented as $\tilde{\Lambda}$ (equivalently referred to as $\epsilon_2$ in \cite{Cardoso:2018ptl}, whose notation we will adopt). 
Within this framework, nonspinning black holes are described by the Schwarzschild metric, and their axial gravitational perturbations yield a modified master equation:
\begin{equation}
    f \frac{d}{dr} 
    \left(f\frac{d\Phi}{dr} \right)+ [\omega^2 -  (V_{\rm RW} + \delta V)] \Phi = 0, \quad 
\end{equation}
where 
$f = 1 - r_H/r$, 
and 
$V_{\rm RW}$ is given by Eq.~\eqref{eq:rwpotential}, while the extra term in the effective potential reads:
\begin{equation}
    \delta V = f \frac{18(\ell + 2)(\ell + 1)\ell(\ell - 1)r_H^8} {r^{10}}\epsilon_2.
\end{equation}
The only coefficient of the expansion Eq.~\eqref{eq:effectivepotential} in Sec.~\ref{sec:pQNMs} is then:
\begin{equation}
    \alpha_{10} = 18(\ell + 2)(\ell + 1)\ell(\ell - 1)\epsilon_2,
\end{equation}
Thus, the quasinormal frequencies are 
expressed as
\begin{align}
\omega &= \omega_{\rm Sch} + \alpha_{10} e_{10}
\notag\\&=
\frac{\Omega_{\rm Sch}}{M} + 18(\ell + 2)(\ell + 1)\ell(\ell - 1)\epsilon_2 e_{10},
\end{align}
where we used that $\omega_{\rm Sch}$ is given by Eq.~\eqref{eq:omega0} and $r_H = 2 M$.
Table.~\ref{table:EFT} shows the quasinormal frequencies for EFT with $\ell= 2,3$. A significant deviation from the predictions of GR in both the real and imaginary parts can be observed in higher overtones.
We report that the same trend is confirmed  
for $\ell=4, 5, 6$ cases.

We note that $\epsilon_2 \lesssim {\cal O}(1)$ in the observational constraints~\cite{Sennett:2019bpc,Silva:2022srr} and theoretical arguments~\cite{Chen:2021bvg,deRham:2021bll,Serra:2022pzl,CarrilloGonzalez:2023cbf,Melville:2024zjq}.
For the validity of the expansion in Eq.~\eqref{eq:pQNMsomega},
Eq.~\eqref{eq:validityofexpansion} should be satisfied.
For example, if we set $\epsilon_2 = 10^{-5}$, Eq.~\eqref{eq:thresholdestimation} holds for $n = 7$ and $\ell = 2$ in Table.~\ref{table:EFT}.
This implies that the expansion in Eq.~\eqref{eq:pQNMsomega} is a good approximation for $n \le 7$ and $\ell = 2$
if $\epsilon_2 \ll 10^{-5}$.

\begin{table}[tp]
\centering
\caption{Quasinormal frequencies in EFT (with $\epsilon_2$ dependency).}
\label{table:EFT}
\begin{subtable}{0.49\textwidth}
\centering
\caption{$\ell=2$}
\begin{tabular}{lc}
\hline
$n$ & $2 M \omega$ \\
\hline
$0$ & $0.74734 - 0.17792i + (1.5921 + 2.8186i) \epsilon_2$ \vspace{-4pt}\\
$1$ & $0.69342 - 0.54783i + (6.2213 + 10.0684i) \epsilon_2$ \vspace{-4pt}\\
$2$ & $0.60211 - 0.95655i + (20.768 + 22.585i) \epsilon_2$ \vspace{-4pt}\\
$3$ & $0.50301 - 1.4103i + (58.262 + 45.413i) \epsilon_2$ \vspace{-4pt}\\
$4$ & $0.41503 - 1.8937i + (147.49 + 87.27i) \epsilon_2$ \vspace{-4pt}\\
$5$ & $0.33860 - 2.3912i + (356.32 + 168.61i) \epsilon_2$ \vspace{-4pt}\\
$6$ & $0.26650 - 2.8958i + (877.52 + 355.27i) \epsilon_2$ \vspace{-4pt}\\
$7$ & $0.18564 - 3.4077i + (2541.0 + 1005.3i) \epsilon_2$ \vspace{-4pt}\\
$8$ & $-4i + (1.6745\times10^5 i) \epsilon_2$ \vspace{-4pt}\\
$9$ & $0.12653 - 4.6053i - (6593.9 - 1682.1i) \epsilon_2$ \vspace{-4pt}\\
$10$ & $0.15311 - 5.1217i - (5171.6 - 826.2i) \epsilon_2$ \\
\hline
\end{tabular}
\end{subtable}
\hfill
\begin{subtable}{0.49\textwidth}
\centering
\caption{$\ell=3$}
\begin{tabular}{lc}
\hline
$n$ & $2 M \omega$ \\
\hline
0 & $1.19889 - 0.18541i + (5.0706 + 5.6488i) \epsilon_2$ \vspace{-4pt}\\
1 & $1.16529 - 0.56260i + (10.336 + 18.267i) \epsilon_2$ \vspace{-4pt}\\
2 & $1.1034 - 0.9582i + (23.240 + 34.872i) \epsilon_2$ \vspace{-4pt}\\
3 & $1.0239 - 1.3807i + (48.491 + 58.146i) \epsilon_2$ \vspace{-4pt}\\
4 & $0.9403 - 1.8313i + (93.205 + 90.609i) \epsilon_2$ \vspace{-4pt}\\
5 & $0.8628 - 2.3043i + (167.27 + 134.50i) \epsilon_2$ \vspace{-4pt}\\
6 & $0.7953 - 2.7918i + (283.77 + 191.88i) \epsilon_2$ \vspace{-4pt}\\
7 & $0.7380 - 3.2877i + (459.23 + 264.76i) \epsilon_2$ \vspace{-4pt}\\
8 & $0.6892 - 3.7881i + (714.04 + 355.10i) \epsilon_2$ \vspace{-4pt}\\
9 & $0.6474 - 4.2908i + (1073.01 + 464.75i) \epsilon_2$ \vspace{-4pt}\\
10 & $0.6109 - 4.7947i + (1566.0 + 595.5i) \epsilon_2$ \\
\hline
\end{tabular}
\end{subtable}
\end{table}

\subsection{Scalar Gauss-Bonnet gravity} 
In this subsection, we will discuss the case of the scalar Gauss-Bonnet (sGB) gravity~\cite{Green:1984sg,Metsaev:1987zx,Antoniadis:1992sa,Mignemi:1992nt,Mignemi:1993ce,Antoniadis:1993jc,Kanti:1995vq, Blazquez-Salcedo:2016enn}.
The sGB gravity is a typical model of modified gravity with a single scalar field, and it is also recognized as a low-energy effective theory that emerges from string theory~\cite{Mignemi:1993ce,Antoniadis:1993jc,Mignemi:1992nt,Kanti:1995vq}.
Adopting the notation from~\cite{Hirano:2024pmk},
the Lagrangian for the shift symmetric sGB gravity is given by
\begin{equation}
    {\cal L} =
    \frac{M_{\rm Pl}^2}{2}R - \frac{1}{2}(\partial \phi)^2
+ \frac{b_1}{\Lambda} \phi \left( R^2 - 4 R_{\mu\nu}R^{\mu\nu} + R_{\mu\nu\rho\sigma}R^{\mu\nu\rho\sigma}\right),
\end{equation}
where $M_{\rm Pl}$ is the Planck mass, $\Lambda$ is the cutoff scale, $b_1$ is the coupling constant. In the case of weak coupling, the sGB term is treated as correction, so we can obtain the contribution of the sGB term to the GR values perturbatively.

Let us consider static spherically symmetric black hole solutions.
The sGB term allows the scalar field to acquire a background value~\cite{Yunes:2011we}. Consequently, the metric changes from the Schwarzschild one. For example, the horizon scale $r_{H}$ departs from the Schwarzschild radius,    
\begin{equation}
    r_{H} = 2M\left(1 -\epsilon^2 b^2_1 \frac{98}{5}\right),
\end{equation}
where $M$ is the ADM mass, and $\epsilon = 1/(4\Lambda M_{\rm Pl}M^2)$ is a small parameter. 
The master equation governing the axial gravitational perturbations around the static spherically symmetric solutions is modified by the sGB correction~\cite{Hirano:2024pmk}, 
which can be expressed as:
\begin{equation}
    f \frac{d}{dr}\left( f
    \frac{d\Phi}{dr}
    \right)  + \left[ \tilde\omega^2 - (V_{\rm RW} + \delta V) \right]\Phi = 0,\label{eq: sGB master eq}
\end{equation}
where $f = 1- r_H/r$ and
\begin{align}
    \tilde\omega^2 &=
    \left(1+\frac{4b_1^2 \epsilon^2}{15}\right)\omega^2,\label{eq: tildeomega}
\\
    \delta V &= f \bigg[\frac{4 b_1^2 \epsilon^2 \omega^2}{15 r^5} \left( r^5 - 146 r^4 r_H - 263 r^3 r_H^2 - 120 r^2 r_H^3 + 98 r r_H^4 + 340 r_H^5 \right) 
\nonumber \\
    &\quad + \frac{b_1^2 \epsilon^2}{15 r^9}  r_H \big\{ 294 (-4 + \ell + \ell^2) r^6 + 87 (-9 + 2 \ell + 2 \ell^2) r^5 r_H
\nonumber \\
    &\quad - 6 (-1825 + 221 \ell + 221 \ell^2) r^4 r_H^2 - (647 + 646 \ell + 646 \ell^2) r^3 r_H^3 
\nonumber \\
    &\quad - 676 (-6 + \ell + \ell^2) r^2 r_H^4 + 10 (-6601 + 218 \ell + 218 \ell^2) r r_H^5 + 56400 r_H^6 \big\}\bigg].\label{eq: sGB potential}
\end{align}
The expansion coefficients in Eq.(\refeq{eq:effectivepotential}) can be obtained by reading off from the potential term (\ref{eq: sGB potential}) up to order $\epsilon^2$: 
\begin{align}
    \alpha_0 &= \frac{4}{15} b_1^2  \epsilon^2 r_H^2\omega^2,\\
    \alpha_1 &= -\frac{584}{15} b_1^2  \epsilon^2 r_H^2\omega^2,\\
    \alpha_2 &= -\frac{1052}{15} b_1^2  \epsilon^2 r_H^2 \omega^2,\\
    \alpha_3 &= \frac{2}{5} b_1^2 \epsilon^2 \left(-196 + 49 \ell + 49 \ell^2 - 80 r_H^2 \omega^2\right), \\
    \alpha_4 &= \frac{1}{15} b_1^2 \epsilon^2 \left(-783 + 174 \ell + 174 \ell^2 + 392 r_H^2 \omega^2\right), \\
    \alpha_5 &= \frac{2}{15} b_1^2 \epsilon^2 \left(5475 - 663 \ell - 663 \ell^2 + 680 r_H^2 \omega^2\right), \\
    \alpha_6 &= -\frac{1}{15} b_1^2 \epsilon^2\left(647 + 646 \ell + 646 \ell^2\right), \\
    \alpha_7 &= -\frac{676}{15} b_1^2 \epsilon^2\left(-6 + \ell + \ell^2\right), \\
    \alpha_8 &= \frac{2}{3} b_1^2 \epsilon^2\left(-6601 + 218 \ell + 218 \ell^2\right), \\
    \alpha_9 &= 3760 b_1^2 \epsilon^2. 
\end{align}
The parametrized QNM formalism is applied to the renormalized $\tilde\omega$ in Eq.~(\ref{eq: tildeomega}). 
The quasinormal frequencies are given by
\begin{align}
    \omega &= \left(1 -\frac{2b_1^2}{15}\epsilon^2\right)\omega_{\rm Sch} +\sum_{j=0}^{9}\alpha_j e_j +{\cal O}(\epsilon^4)
    \notag\\&= \frac{\Omega_{\rm Sch}}{M} +
    \frac{b_1^2 \epsilon^2}{15 M}
\bigg[
292 \Omega_{\rm Sch} 
+16 M\Omega_{\rm Sch}^2 (e_0-146 e_1-263 e_2-120 e_3+98 e_4+340 e_5) 
\notag\\&\quad 
+ 
M \Big( -783 e_4+10950 e_5-647 e_6+4056 e_7-66010 e_8+56400 e_9 + 294 e_3 (\ell^2+\ell-4)
\notag\\&\quad  
       + 2 \ell (\ell+1) (87 e_4-663 e_5-323 e_6-338 e_7+1090 e_8)
   \Big)
\bigg] +{\cal O}(\epsilon^4).
\end{align}
The resultant quasinormal frequencies for $\ell=2$ and $\ell=3$ are in Table.~\ref{fig: sGB table}. 
The imaginary part of the quasinormal frequencies in the sGB gravity has a growing trend as the overtone number becomes higher.

The observational constraints~\cite{Sennett:2019bpc,Silva:2022srr} and theoretical arguments~\cite{Chen:2021bvg,deRham:2021bll,Serra:2022pzl,CarrilloGonzalez:2023cbf,Melville:2024zjq} imply $b_1^2 \epsilon^2 \lesssim {\cal O}(1)$.
For the expansion in~Eq.~\eqref{eq:pQNMsomega} to be valid, Eq.~\eqref{eq:validityofexpansion} should be satisfied.
For example, considering $b_1^2 \epsilon^2 = 10^{-2}$,
Eq.~\eqref{eq:thresholdestimation} holds for $n = 10$ and $\ell = 2$ in Table.~\ref{fig: sGB table}.
Therefore, if $b_1^2 \epsilon^2 \ll 10^{-2}$, the expansion in Eq.~\eqref{eq:pQNMsomega} is a good approximation for $n \le 10$ and $\ell = 2$.

\begin{table}[tp]
\centering
\caption{Quasinormal frequencies in sGB gravity for various $\ell$ values.}\label{fig: sGB table}
\begin{subtable}{0.49\textwidth}
\centering
\caption{$\ell=2$}
\begin{tabular}{lc}
\hline
$n$ & $2 M \omega$\\
\hline
0 & $0.74734 - 0.17792i + (14.548 - 3.464i) b_1^2 \epsilon^2$ \vspace{-3pt} \\
1 & $0.69342 - 0.54783i + (13.499 - 10.664i) b_1^2 \epsilon^2$ \vspace{-3pt} \\
2 & $0.60211 - 0.95655i + (11.721 - 18.621i) b_1^2 \epsilon^2$\vspace{-3pt} \\
3 & $0.5030 - 1.4103i + (9.792 - 27.454i) b_1^2 \epsilon^2$ \vspace{-3pt}\\
4 & $0.4150 - 1.8937i + (8.079 - 36.864i) b_1^2 \epsilon^2$ \vspace{-3pt}\\
5 & $0.3386 - 2.3912i + (6.591 - 46.549i) b_1^2 \epsilon^2$ \vspace{-3pt}\\
6 & $0.2665 - 2.8958i + (5.188 - 56.372i) b_1^2 \epsilon^2$ \vspace{-3pt}\\
7 & $0.1856 - 3.4077i + (3.614 - 66.336i) b_1^2 \epsilon^2$ \vspace{-3pt}\\
8 &  $ -4i - (77.867 i) b_1^2 \epsilon^2$         \vspace{-3pt}\\
9 & $0.1265 - 4.6053i + (2.463 - 89.650i) b_1^2 \epsilon^2$\vspace{-3pt} \\
10 & $0.1531 - 5.1217i + (2.980 - 99.702i) b_1^2 \epsilon^2$ \\
\hline
\end{tabular}
\end{subtable}
\hfill
\begin{subtable}{0.49\textwidth}
\centering
\caption{$\ell=3$}
\begin{tabular}{lc}
\hline
$n$ & $2 M \omega$ \\
\hline
0 & $1.19889 - 0.18541i + (23.338 - 3.609i) b_1^2 \epsilon^2$ \vspace{-3pt}\\
1 & $1.16529 - 0.56260i + (22.684 - 10.952i) b_1^2 \epsilon^2$ \vspace{-3pt}\\
2 & $1.1034 - 0.9582i + (21.479 - 18.653i) b_1^2 \epsilon^2$ \vspace{-3pt}\\
3 & $1.0239 - 1.3807i + (19.932 - 26.877i) b_1^2 \epsilon^2$ \vspace{-3pt}\\
4 & $0.9403 - 1.8313i + (18.305 - 35.649i) b_1^2 \epsilon^2$ \vspace{-3pt}\\
5 & $0.8628 - 2.3043i + (16.795 - 44.857i) b_1^2 \epsilon^2$ \vspace{-3pt}\\
6 & $0.7953 - 2.7918i + (15.482 - 54.348i) b_1^2 \epsilon^2$ \vspace{-3pt}\\
7 & $0.7380 - 3.2877i + (14.366 - 64.000i) b_1^2 \epsilon^2$ \vspace{-3pt}\\
8 & $0.6892 - 3.7881i + (13.417 - 73.741i) b_1^2 \epsilon^2$ \vspace{-3pt}\\
9 & $0.6474 - 4.2908i + (12.602 - 83.528i) b_1^2 \epsilon^2$ \vspace{-3pt}\\
10 & $0.6109 - 4.7947i + (11.893 - 93.337i) b_1^2 \epsilon^2$ \\
\hline
\end{tabular}
\end{subtable}
\end{table}
\subsection{Slowly rotating Kerr black holes}

In this section, we present a comparison between numerical results~\cite{Cook:2014cta, Cook:2023data} and the first-order semi-analytic expression of QNMs for slowly rotating Kerr black holes, as we briefly demonstrate how the parametrized QNM formalism can be used to derive the semi-analytic QNM frequency of the Kerr black hole.

\begin{figure}[tbp]
\begin{center}
\includegraphics[width=0.31\linewidth]{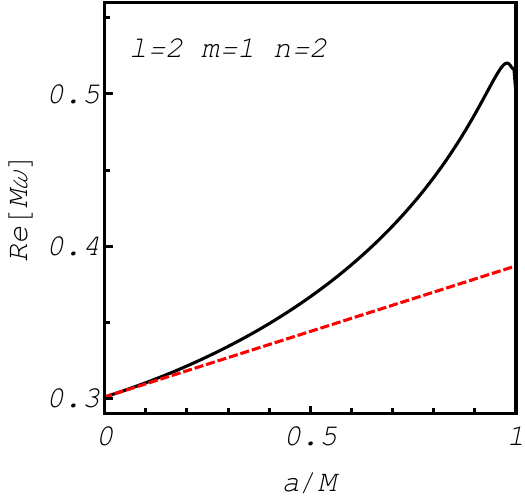}
\quad
\includegraphics[width=0.31\linewidth]{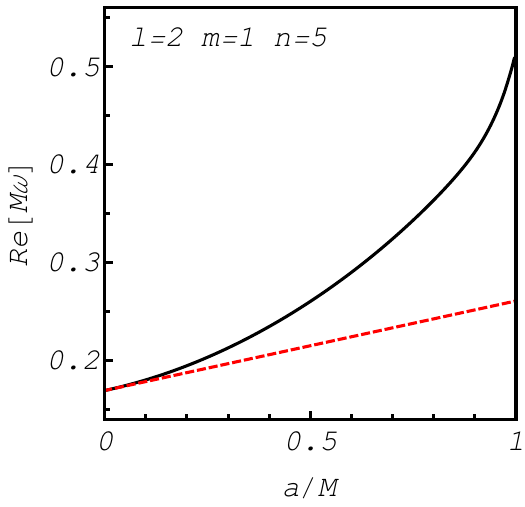}
\quad
\includegraphics[width=0.31\linewidth]{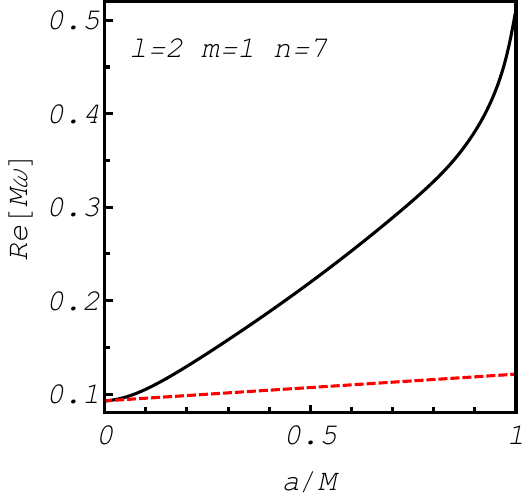}
\includegraphics[width=0.31\linewidth]{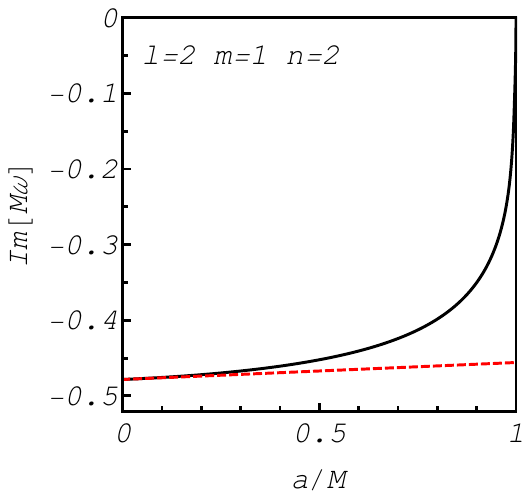}
\quad
\includegraphics[width=0.31\linewidth]{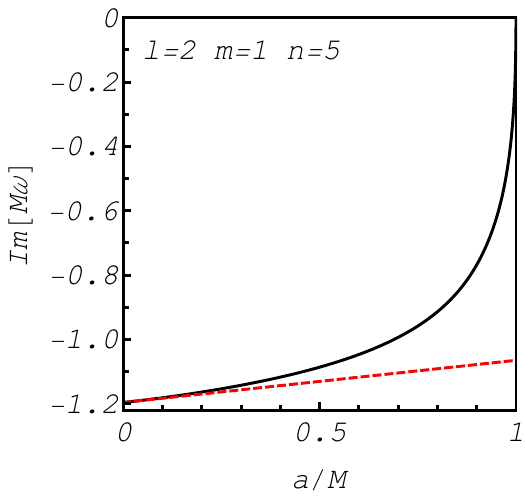}
\quad
\includegraphics[width=0.31\linewidth]{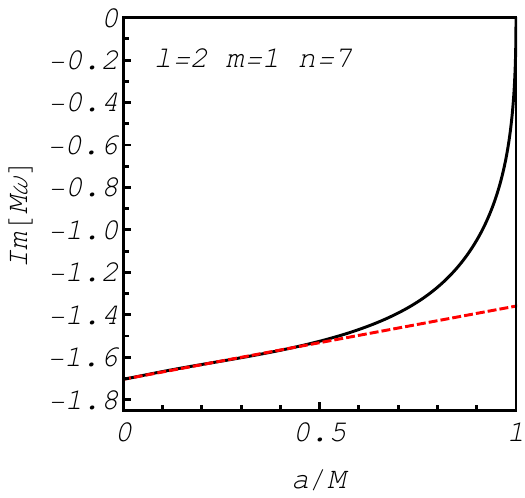}
\caption{
Comparison with numerical calculation and our formula.
Black lines denote the numerical result in~\cite{Cook:2014cta, Cook:2023data} and 
red dashed lines denote our formula for slowly rotating Kerr black holes~\eqref{eq:qnmslowkerr}.
We plotted cases for $\ell =2, m =1$ and overtone numbers $n=2, 5, 7$.
}
\label{fig:slowrotkerrqnm}
\end{center}
\end{figure}

\begin{figure}[tbp]
\begin{center}
\includegraphics[width=0.31\linewidth]{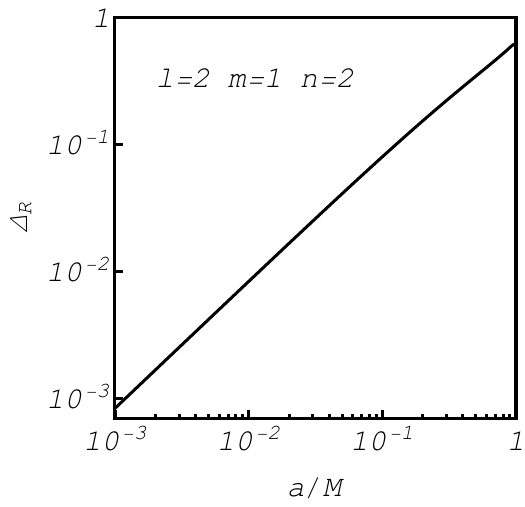}
\quad
\includegraphics[width=0.31\linewidth]{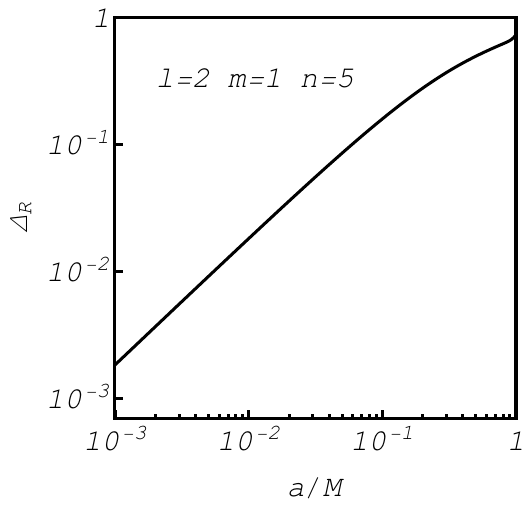}
\quad
\includegraphics[width=0.31\linewidth]{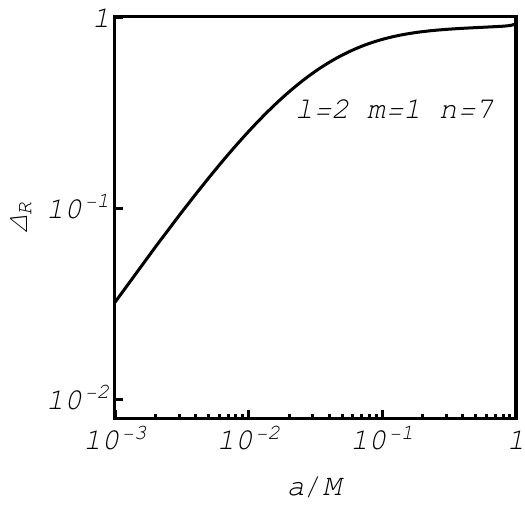}
\includegraphics[width=0.31\linewidth]{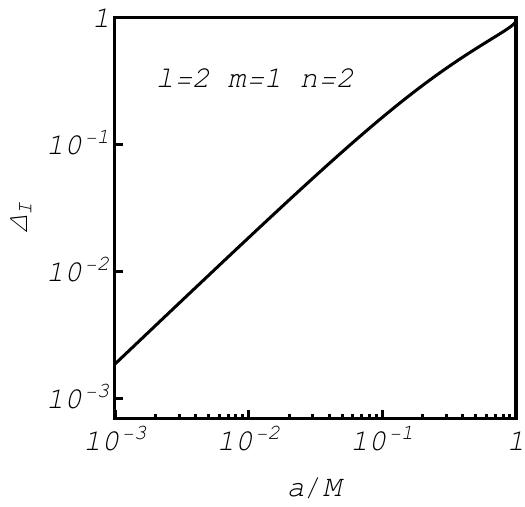}
\quad
\includegraphics[width=0.31\linewidth]{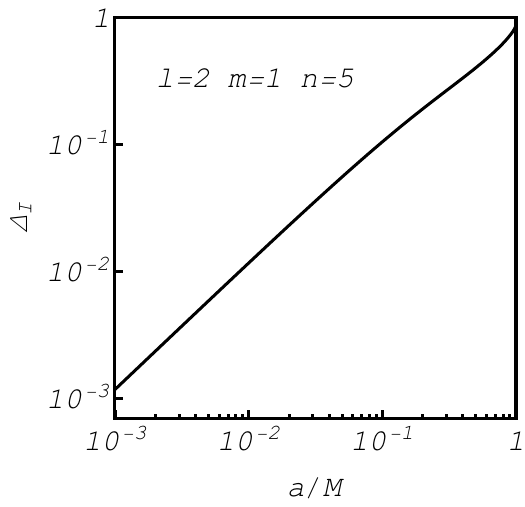}
\quad
\includegraphics[width=0.31\linewidth]{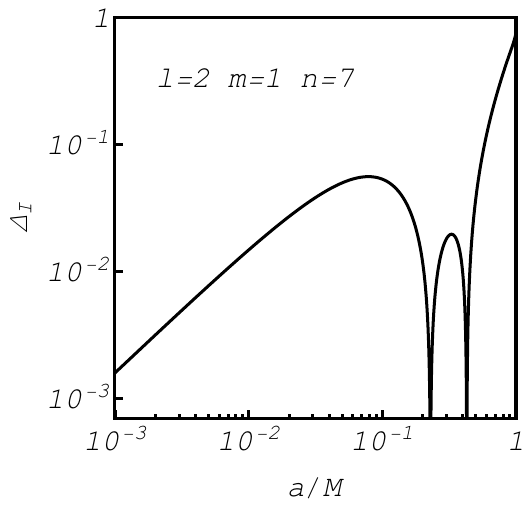}
\caption{Analysis of error functions in Eq.~\eqref{eq:kerr_error_re} and \eqref{eq:kerr_error_im} corresponding to modes with $\ell = 2$, $m = 1$, and overtone numbers $n = 2, 5, 7$.
}
\label{fig:slowrotkerrerror}
\end{center}
\end{figure}

The Chandrasekar-Detweiler equation~\cite{Chandrasekhar:1976zz, Detweiler:1977gy}, 
derived from the Teukolsky equation~\cite{Teukolsky:1973ha} through variable transformations, reduces to the Regge-Wheeler equation in the Schwarzschild limit.
As detailed in \cite{Hatsuda:2020egs}, 
for slowly rotating cases,
the Chandrasekar-Detweiler equation
can be rewritten in the form 
such that the parametrized QNM formalism 
is applicable
\begin{equation}
    f\frac{d}{dr}\left( f\frac{d}{dr}\right)\Phi + \bigg( \left(\omega-\frac{m}{r_H}\sqrt{\frac{r_{-}}{r_H}} \right)^2 -(V_{\rm RW}+\delta V) \bigg)\Phi=0,
\end{equation}
where 
$f = 1-r_H/r, r_H = M + \sqrt{M^2 - a^2}, r_- = M - \sqrt{M^2 - a^2}$
and $\delta V$ denotes the deviation from the Regge-Wheeler potential.
At the first order of $\sqrt{r_-/r_H} \simeq a/(2M)$, the deviation $\delta V$ is given by
\begin{align}
    \delta V =\frac{f}{r_H^2} \sum_{j=0}^{5} \alpha_j \left( \frac{r_H}{r} \right)^j, 
\end{align}
where the coefficients 
$\alpha_j$ are given by
\begin{align}
\alpha_0 &= -2m\omega_{\rm Sch} r_H \sqrt{\frac{r_-}{r_H}},\\
\alpha_1 &= -2m\omega_{\rm Sch} r_H \sqrt{\frac{r_-}{r_H}},\\
\alpha_2 &= m\omega_{\rm Sch} r_H 
\left(-2 - \frac{8}{\ell (\ell + 1)}\right)
\sqrt{\frac{r_-}{r_H}},\\
\alpha_3 &= \frac{8m(3 + 2(\ell^2 + \ell - 2))}{3\omega_{\rm Sch} r_H} \sqrt{\frac{r_-}{r_H}},\\
\alpha_4 &= -\frac{4m(5 + 2(\ell^2 + \ell - 2))}{\omega_{\rm Sch} r_H} \sqrt{\frac{r_-}{r_H}},\\
\alpha_5 &= \frac{12m}{\omega_{\rm Sch} r_H} \sqrt{\frac{r_-}{r_H}}.
\end{align}
Hence, the QNM frequency for slowly rotating Kerr black hole is given by
\begin{align}
\omega &= \omega_{\rm Kerr}^{\rm 1st} 
=
\omega_{\rm Sch}
+\frac{m}{r_H}
\sqrt{\frac{r_{-}}{r_H}}  
+\sum_{j=0}^{5} \alpha_j e_j,
\notag \\
&= 
\frac{\Omega_{\rm Sch}}{M}
+
\bigg[
\frac{1}{4}
-
2 M \Omega_{\rm Sch} e_0
-
2 M \Omega_{\rm Sch} e_1
-
 \bigg(2 + \frac{8}{\ell + \ell^2} \bigg) M\Omega_{\rm Sch} e_2
\notag \\& \quad 
+
\frac{-2 + 4 \ell(\ell + 1)}{3\Omega_{\rm Sch}} M e_3
-
\frac{1 + 2 \ell (\ell + 1)}{\Omega_{\rm Sch}} M e_4
+
\frac{3 M e_5}{\Omega_{\rm Sch}} 
\bigg] \frac{m a}{M^2}
+ {\cal O}(a^2).
\label{eq:qnmslowkerr}
\end{align}
In Fig.~\ref{fig:slowrotkerrqnm}, we plot the quasinormal frequencies for our first order formula~\eqref{eq:qnmslowkerr}
and the numerical result in~\cite{Cook:2014cta, Cook:2023data}.
In Fig.~\ref{fig:slowrotkerrerror}, we also plot 
the error functions defined by
\begin{align}
\Delta_R &= 
\left| \frac{{\rm Re}[(\omega_{\rm full} - \omega_{\rm Sch}) - (\omega_{\rm Kerr}^{\rm 1st} - \omega_{\rm Sch})]}{{\rm Re}[\omega_{\rm full} - \omega_{\rm Sch}]} \right|,
\label{eq:kerr_error_re}
\\
\Delta_I &= 
\left| \frac{{\rm Im}[(\omega_{\rm full} - \omega_{\rm Sch}) - (\omega_{\rm Kerr}^{\rm 1st} - \omega_{\rm Sch})]}{{\rm Im}[\omega_{\rm full} - \omega_{\rm Sch}]} \right|,
\label{eq:kerr_error_im}
\end{align}
where $\omega_{\rm full}$ denotes the numerical result in~\cite{Cook:2014cta, Cook:2023data} and 
$\omega_{\rm Kerr}^{\rm 1st}$ denotes our first order formula~\eqref{eq:qnmslowkerr}.
Across various overtones, the imaginary part of the expression functions satisfactorily; however, the real part increasingly departs from numerical calculations as the overtone number $n$ rises.

\section{Summary and Discussion}

In this paper, we studied the parametrized QNMs formalism, especially, paying attention to overtones.
In this formalism, minor modifications to the GR case effective potential are expressed in series expansion forms, and they modify QNM behaviors at linear order in the deviations from GR.
We adopted Nollert's method, known for its improved control over the larger imaginary parts, to numerically compute the precise values of 
the coefficients $e_j$.
From the results, we observed that in contrast to the fundamental mode, there exists a growing trend in the coefficients $e_j$ as $j$ increases, and this trend is more evident as the overtone number $n$ goes up.
This suggests that the correction to the QNM frequency contributes significantly to overtones at large $j$.
In typical modified gravity theories with higher curvature correction terms, we have confirmed that the deviation of the quasinormal frequencies from the GR value grows for higher overtones.
Our findings indicate that if the overtones of QNMs are observed, they could be used to put a strong constraint on the gravity theories.

The growing tendency of $e_j$ for overtones can also be understood using an analytical method.
From the recursion relation for $e_j$ in Eq.~\eqref{eq:ejrecrelation},
we derived the large $j$ asymptotic behavior as $e_j \simeq C j^{-1 - 2r_H \omega_I} e^{2ir_H \omega_R \ln j}$, where $\omega_{\rm Sch} = \omega_R + i\omega_I$.
This implies that the magnitude of $e_j$ is characterized by the factor
$j^{-1 - 2r_H \omega_I}$. We confirmed that 
the power $-1 - 2r_H \omega_I$ becomes positive for overtones where $\omega_I$ has negative larger values than the fundamental mode.
This is the essential reason why the magnitude of $e_j$ for higher overtones grows 
as $j$ increases.
Also, since
$\delta V_j$ in Eq.~\eqref{eq:effectivepotential} 
has a peak near the horizon $r_* = - r_H \ln(j/e)$ for large $j$,
the increasing trend of the coefficient $e_j$ for large $j$ implies that modifications to the effective potential in the near-horizon regime strongly influence the overtones of QNMs, 
which is consistent with recent discussion in~\cite{Konoplya:2022pbc}.
Our result also helps us to understand recently confirmed findings that the overtone spectra are relatively more unstable against perturbations in the effective potential when compared to the fundamental mode~\cite{Jaramillo:2020tuu, Jaramillo:2021tmt, Cheung:2021bol, Berti:2022xfj}.
Since higher curvature modified gravity theories 
typically have modification terms with large $j$ in the effective potential as discussed in Sec~\ref{sec:applications},
the spectral instability of overtones is realized as the growing trend of quasinormal frequencies for overtones in such gravity theories.
Our results are another physically important example of the QNM spectral instability~\cite{Cardoso:2024mrw}.

Finally, we comment on the caveats of the extraction of the overtones from ringdown data.
Considering the rapidly decaying nature of the overtones, their extractions are expected to be performed at the early ringdown phase with great care. 
The competing effects of non-linear and linear gravitational perturbations should also be taken into account, as the contribution of nonlinearity to the early ringdown phase may be significant.
Many recent studies have been tackling linear~\cite{Baibhav:2017jhs,
London:2014cma,Giesler:2019uxc,Bhagwat:2019dtm,CalderonBustillo:2020rmh,Forteza:2021wfq,Dhani:2021vac,Finch:2021iip,Finch:2021qph,MaganaZertuche:2021syq,Ota:2019bzl,JimenezForteza:2020cve,Mourier:2020mwa,Cook:2020otn,Ota:2021ypb,Dhani:2020nik} and non-linear analysis~\cite{Gleiser:1995gx, Campanelli:1998jv, Zlochower:2003yh, Ioka:2007ak, Nakano:2007cj, Brizuela:2009qd, Pazos:2010xf, Loutrel:2020wbw, Ripley:2020xby, Sberna:2021eui, Ma:2022wpv, Mitman:2022qdl, Cheung:2022rbm, Khera:2023oyf, Lagos:2022otp, Guerreiro:2023gdy, Kehagias:2023ctr, Kehagias:2023mcl, Redondo-Yuste:2023seq, Perrone:2023jzq, Bucciotti:2023ets, Cheung:2023vki, Ma:2024qcv, Yi:2024elj, Okounkova:2020vwu}. 
At the same time, we need to care about the determination of the starting time for the ringdown phase and the possibility of overfitting \cite{Nee:2023osy,Baibhav:2023clw,Franchini:2023eda}.
Toward the detection of overtones in real GW data, we need an enhanced approach to the aforementioned issues in addition to the detector sensitivity.

\acknowledgments{
We would like to thank Emanuele Berti, Yasuyuki Hatsuda and Hiroyuki Nakano for useful comments and discussions. 
This research is supported by JSPS KAKENHI Grant Nos. JP22K03626(M.K.) and JP21H01080(M.Y.), and IBS under the project code IBS-R018-D3 (M.Y. and J.Z.).
}

\appendix

\section{Recursion relations among the coefficients $e_j$}
\label{appendixejrecrel}

When analyzing parametrized QNMs, one can uncover a recursion relation (\cite{Kimura:2020mrh}, \cite{Hatsuda:2023geo}) between the coefficients $e_j$ with different subscripts $j$.
This relation demonstrates the correlation between consecutive terms in the expansion. For the odd-parity case that we study, it is defined as Eq.~\eqref{eq:ejrecrelation}.
\begin{equation}
    c_{j+1} e_{j+1}+ c_{j+3} e_{j+3} + c_{j+4} e_{j+4} + c_{j+5}e_{j+5} = 0,
   \label{eq:ejrecrelation}
\end{equation}
where the explicit form of $c_{j+a}$ can be expressed as
\begin{align}
c_{j+1} &= -4j r_H^2 (\omega_{\rm Sch})^2, 
\label{eq:cj1}
\\
c_{j+3} &= -(\ell(\ell + 1) - j(j + 2)), \\
c_{j+4} &= (2j + 3)(-6 - 2\ell(\ell + 1) + j(j + 3)), \\
c_{j+5} &= -(j - 2)(j + 2)(j + 6),  
\label{eq:cj5}
\end{align}
Such recursive structures are a result of perturbative methods applied to the gravitational equations governing the dynamics of the system under study as well as underlying symmetries.
The determination of the coefficients $e_j$ for the parametrized QNM formalism can only be done manually, as the underlying equations are excessively complex. 
The identified recursive relations provide a notable computational benefit, enabling the estimation of coefficients for higher $j$ via only the numerically computed values 
of $e_j$ terms for lower $j$.

\section{Non-trivial relation including $e_1$}
\label{appendix:e1}
It should be noted that the recursion relation described in the previous section does not include the relation between the coefficient $e_1$ and the other components. Nevertheless, this non-trivial connection can be obtained by following the subsequent strategy:

In~\cite{Hatsuda:2020iql}, 
the equation 
\begin{align}
\left(1-\frac{r_H}{r}\right) \frac{d}{dr}\left(\left(1-\frac{r_H}{r}\right)\frac{d\tilde{\Phi}}{dr}\right)
+
((2 M \omega)^2 - \tilde{V}) \tilde{\Phi} = 0,
\end{align}
with 
\begin{align}
\tilde{V} = \left(1-\frac{r_H}{r}\right)
\bigg(
\frac{4 a^2 \omega^2}{r_H^2}
+
\frac{4 a \omega (m - a \omega)}{r_H r}
+
\frac{A_{\ell,m} + 2 - a \omega(2 m - a \omega)}{r^2}
- 
\frac{3 r_H}{r^3}
\bigg),
\end{align}
is proposed as an isospectral equation 
to the radial spin-$2$ Teukolsky equation~\cite{Teukolsky:1973ha}.
Taking the slow rotation limit, we can rewrite the potential as
\begin{align}
 \tilde{V} &= 
 \left(1-\frac{r_H}{r}\right)
 \left(
 \frac{\ell(\ell + 1)}{r^2} - \frac{3r_H}{r^3}
 + \frac{\tilde{\alpha}_1}{r_H^2} \left(\frac{r_H}{r}\right)
 + \frac{\tilde{\alpha}_2}{r_H^2} \left(\frac{r_H}{r}\right)^2
 +
 {\cal O}(a^2)
 \right),
\end{align}
with
\begin{align}
\tilde{\alpha}_1 &= 4 m r_H \omega_{\rm Sch} \sqrt{\frac{r_-}{r_H}},
\\
\tilde{\alpha}_2 &= -2\left(1 + \frac{4}{\ell + \ell^2}\right) 
  m r_H \omega_{\rm Sch}
  \sqrt{\frac{r_-}{r_H}}.
\end{align}
Using the parametrized QNM formalism,
we obtain the QNM frequency for a slowly rotating Kerr black hole as
\begin{align}
\omega = \omega_{\rm Sch} + \tilde{\alpha}_1 e_1 + \tilde{\alpha}_1 e_2.
\end{align}
Comparing this formula to Eq.~\eqref{eq:qnmslowkerr}, we obtain a non-trivial relation which contains $e_1$ as
\begin{align}
-3 \omega_{\rm Sch}
+
6 r_H^2 \omega_{\rm Sch}^2 e_0
+
18 r_H^2 \omega_{\rm Sch}^2 e_1
-
8 (-1 + 2 \ell + 2 \ell^2) e_3
+
12(1 + 2 \ell + 2 \ell^2) e_4 
- 
36 e_ 5 = 0.
\end{align}
Using Eq.~\eqref{eq:ejrecrelation},
this relation can be rewritten as a simpler form
\begin{align}
-6 \omega_{\rm Sch} + 36 r_H^2 \omega_{\rm Sch}^2 e_1
- 2 (-20 + 7 \ell + 7 \ell^2) e_3
+
15 (-5 + 2 \ell + 2 \ell^2) e_4 = 0.
\end{align}

\section{Coefficients $e_j$ for $n=8, \ell = 2$ overtone}
\label{sec:l2n8}

In the Schwarzschild limit, there are special solutions where the series in Eq.~\eqref{eq:frobeniusseries} becomes a finite series, making $\sum_{k =0}^\infty a_k$ clearly convergent.
Specifically, for the $\ell = 2$ case, the frequency for such solution is given by $\omega = -4i/r_H$ and the wave function becomes
\begin{align}
\Phi_{\rm Sch}^{n=8} &=
C \exp(i\omega r_*) f^\rho
\bigg(
-\frac{945}{4194304}
+\frac{15525 f }{4194304}
-\frac{29025 f ^2}{1048576}
+\frac{130185 f ^3}{1048576}
-\frac{775935 f ^4}{2097152}
\notag\\&\quad
+\frac{1604979 f ^5}{2097152}
-\frac{1164765 f ^6}{1048576}
+\frac{1164765 f ^7}{1048576}
-\frac{2717785 f ^8}{4194304}
+ \frac{1164765 f ^9}{4194304}
\bigg),
\label{eq:assolutionsch}
\end{align}
with $\rho = -2ir_H \omega$.
This solution is known as the algebraically special mode~\cite{Chandrasekhar:1984},
and the frequency $\omega = -4i/r_H$ corresponds to $n = 8$ overtone for $\ell = 2$ Schwarzschild case.
However, we should note that 
according to~\cite{MaassenvandenBrink:2000iwh} there exists an argument that for such special frequency, the QNM boundary condition fails to hold at the horizon.
Nevertheless, in this section, we discuss the shift of frequency of $n =8$ mode for the parametrized potential in Eq.~\eqref{eq:effectivepotential}.

To discuss the coefficient $e_j$ in Eq~\eqref{eq:pQNMsomega} for $n = 8, \ell = 2$ overtone, we assume the same form of the wave function $\Phi$ in Eq.~\eqref{eq:frobeniusseries} 
with $\rho = -2ir_H \omega$
whose GR limit is given by 
Eq.~\eqref{eq:assolutionsch}.
For the parametrized potential in Eq.~\eqref{eq:effectivepotential},
we obtain a recursion relation for $a_j$.
After analyzing this recursion relation, we find the frequency in the form 
\begin{align}
\omega &=  -\frac{4i}{r_H}+ \sum_j \alpha_j e_j,
\end{align}
and the coefficients $e_j$ are given by
\begin{align}
r_H e_0 &= \frac{1164765 i}{700009},
\\
r_H e_1 &= \frac{5571603 i}{1400018},
\\
r_H e_2 &= \frac{5823825 i}{700009},
\\
r_H e_3 &= \frac{44416947 i}{2800036},
\\
r_H e_4 &= \frac{19775313 i}{700009},
\\
r_H e_5 &= \frac{66780435 i}{1400018},
\\
r_H e_6 &= \frac{3176547 i}{41177},
\\
r_H e_7 &= \frac{674140005 i}{5600072}.
\end{align}
Note that these coefficients $e_j$
satisfy the relation in Eq.~\eqref{eq:ejrecrelation} and 
$e_j$ with $j \ge 8$ can be generated from Eq.~\eqref{eq:ejrecrelation}.

Substituting the above coefficients $e_j$ to the formula for slowly rotating Kerr black hole in Eq.~\eqref{eq:qnmslowkerr}, we obtain 
\begin{align}
2 M \omega = -4 i - \frac{33078176}{700009} \frac{m a}{2 M}
+ {\cal O}(a^2).
\label{eq:qnmslowrotkerrn8}
\end{align}
This completely coincides with the result at the order of ${\cal O}(a)$ in~\cite{MaassenvandenBrink:2000iwh, Berti:2003jh, Berti:2009kk}.
In~\cite{Cook:2016fge, Cook:2016ngj}, 
there is a report that
QNM frequencies for slowly rotating Kerr black holes calculated numerically agree with this prediction Eq.~\eqref{eq:qnmslowrotkerrn8} for branches where
$2 M \omega \to -4i$ in $a \to 0$ limit.

\newpage

\section{Dimensionless Schwarzschild quasinormal frequencies $\Omega_{\rm Sch}$ }
\label{sec:omegadata}

We provide numerical values of $\Omega_{\rm Sch}$, which are dimensionless quasinormal frequencies 
normalized by the mass (not the black hole radius) 
for the Schwarzschild black holes in Tables \ref{tab:omega_l2} and \ref{tab:omega_l3}, where $n$ denotes the overtone number.

\begin{table}[H]
\caption{
$\ell=2$ Schwarzschild QNM frequencies normalized by the mass.}
\centering
\begin{tabular}{c c}
\hline
$n$ & $\Omega_{\rm Sch}$ for $\ell=2$ \\ \hline
0 & $0.373671684418041835793492002977 - 0.088962315688935698280460927185i$ \vspace{-4pt}\\
1 & $0.346710996879163439717675359733 - 0.273914875291234817349560222138i$ \vspace{-4pt}\\
2 & $0.301053454612366393802003608880 - 0.478276983223071809984182830723i$ \vspace{-4pt}\\
3 & $0.251504962185590589455540447872 - 0.705148202433495368586197050439i$ \vspace{-4pt}\\
4 & $0.207514579813065579533068679253 - 0.946844890866351547285362117782i$ \vspace{-4pt}\\
5 & $0.169299403093043660543806469280 - 1.195608054135846807847460029672i$ \vspace{-4pt}\\
6 & $0.13325234024518802334878268918 - 1.44791062616203815799480739742i$ \vspace{-4pt}\\
7 & $0.09282233367020094511758196991 - 1.70384117220613552700100693606i$ \vspace{-4pt}\\ 
8  & $-2i $\vspace{-4pt}\\
9 & $0.06326350512560599307041854974  - 2.30264476515854049612775361250i$ \vspace{-4pt}\\
10& $0.07655346288598616222699244005  - 2.56082661738150570057182167367i$  \\ 
\hline
\end{tabular}
\label{tab:omega_l2}
\end{table}

\begin{table}[H]
\caption{
$\ell=3$ Schwarzschild QNM frequencies normalized by the mass.}
\centering
\begin{tabular}{c c}
\hline
$n$ & $\Omega_{\rm Sch}$ for $\ell=3$ \\ \hline
0  & $0.599443288437490072739493977601 - 0.092703047944947603970074106711i$\vspace{-5pt} \\
1  & $0.582643803033299430615840785500 - 0.281298113435044044678838264497i$ \vspace{-4pt}\\
2  & $0.551684900778451316386411789182 - 0.479092750966962234968187545681i$ \vspace{-4pt}\\
3  & $0.511961911058337382119256601539 - 0.690337095969239106511366065005i$ \vspace{-4pt}\\
4  & $0.470174005815155228143427478927 - 0.915649392505096606637106363461i$ \vspace{-4pt}\\
5  & $0.431386478642153844922167144935 - 1.152151362140905438995312994801i$ \vspace{-4pt}\\
6  & $0.39765952417575710857715463748  - 1.39591224272259156879382199256i$ \vspace{-4pt}\\
7  & $0.36899227588972845769529171517  - 1.64384452835676619241220936938i$ \vspace{-4pt}\\
8  & $0.34461831859511639569908026946  - 1.89403280419296282019417602374i$ \vspace{-4pt}\\
9  & $0.32368313160134644137286356638  - 2.14539894976732050705523618001i$ \vspace{-4pt}\\
10 & $0.30546090193109574672725993056  - 2.39735455030161310920191229124i$ \\ \hline
\end{tabular}
\label{tab:omega_l3}
\end{table}


\newpage

\section{Numerical data for $e_j$}
\label{sec:dataej}

We provide numerical values of 
the coefficients $e_j$ in the parametrized QNM formalism for $\ell = 2, 3$ and $n = 0,1,\ldots 10$
in Tables \ref{tab:n0ell23}-\ref{tab:n10ell23}, where the values are normalized by the horizon radius $r_H$.
We note that the coefficients $e_j$ with high $j$ can be generated from the recursion relation~\eqref{eq:ejrecrelation}.

\begin{table}[H]
\caption{$e_j$ data for $n = 0$.}
\vspace{-7pt}
\hspace*{-20pt}
\centering
\begin{subtable}{0.49\textwidth}
\caption{$\ell=2$}
\vspace{-5pt}
\begin{tabular}{c@{\hspace{10pt}}c@{\hspace{10pt}}c}
\hline
$\ell$ & $j$ &  $r_H e_j$ \\
\hline
   & $0$ & $0.247251965436735 + 0.092643073755839i$ \vspace{-4pt}\\
   & $1$ & $0.159854787039703 + 0.018208481773442i$ \vspace{-4pt}\\
   & $2$ & $0.0966322401342324 - 0.0024154964534615i$ \vspace{-4pt}\\
   & $3$ & $0.0584907850084892 - 0.0037178612876936i$ \vspace{-4pt}\\
   & $4$ & $0.0366794367818616 - 0.0004386969534051i$ \vspace{-4pt}\\
$2$& $5$ & $0.0240379477506994 + 0.0027307931425322i$ \vspace{-4pt}\\
   & $6$ & $0.0163428109579658 + 0.0048426716837515i$ \vspace{-4pt}\\
   & $7$ & $0.01136357508132363 + 0.00601399193209594i$ \vspace{-4pt}\\
   & $8$ & $0.00795199773524588 + 0.00653699645743769i$ \vspace{-4pt}\\
   & $9$ & $0.00550386682041598 + 0.00665277086108331i$\vspace{-4pt}\\
   & $10$ & $0.00368531438581036 + 0.00652444451724573i$ \\
\hline
\end{tabular}
\end{subtable}
\hfill
\begin{subtable}{0.49\textwidth}
\caption{$\ell=3$}
\vspace{-5pt}
\begin{tabular}{c@{\hspace{10pt}}c@{\hspace{10pt}}c}
\hline
$\ell$ & $j$ &  $r_H e_j$ \\
\hline
   & $0$ & $0.144427428250107 + 0.036770321741985i$ \vspace{-4pt}\\
   & $1$ & $0.0957683223248566 + 0.0086035480579422i$ \vspace{-4pt}\\
   & $2$ & $0.0614725008272383 - 0.0006195234760257i$ \vspace{-4pt}\\
   & $3$ & $0.0392928459198075 - 0.0020278784624223i$ \vspace{-4pt}\\
   & $4$ & $0.0254322953286445 - 0.0009608457364406i$ \vspace{-4pt}\\
$3$& $5$ & $0.0167854156428019 + 0.0004525854131922i$ \vspace{-4pt}\\
   & $6$ & $0.01128895497550895 + 0.00154867984591634i$ \vspace{-4pt}\\
   & $7$ & $0.00769038142142765 + 0.00222885207817423i$ \vspace{-4pt}\\
   & $8$ & $0.00525650581974478 + 0.00256801534497293i$ \vspace{-4pt}\\
   & $9$ & $0.00355997381645353 + 0.00266834227449026i$ \vspace{-4pt}\\
   & $10$ & $0.00234751849358300 + 0.00261518583054406i$ \\
\hline
\end{tabular}
\end{subtable}%

\label{tab:n0ell23}
\end{table}


\begin{table}[H]
    
\caption{$e_j$ data for $n=1$.}
\vspace{-7pt}
\hspace*{-10pt}
\centering
\begin{subtable}{0.49\textwidth}
\caption{$\ell=2$}
\vspace{-5pt}
\begin{tabular}{c@{\hspace{10pt}}c@{\hspace{10pt}}c}
\hline
$\ell$ & $j$ &  $r_H e_j$ \\
\hline
   & $0$ & $0.107464723514924 + 0.185387391385682i$\vspace{-4pt}\\
   & $1$ & $0.146179872594831 + 0.052434921070417i$ \vspace{-4pt}\\
   & $2$ & $0.1041379948373251 - 0.0044390476970642i$ \vspace{-4pt}\\
   & $3$ & $0.0652386112046042 - 0.0101870692777796i$ \vspace{-4pt}\\
   & $4$ & $0.0442462688718310 - 0.0007443024387448i$ \vspace{-4pt}\\
$2$& $5$ & $0.0343154937570156 + 0.0085123887906039i$ \vspace{-4pt}\\
   & $6$ & $0.0288347045024162 + 0.0146400895953562i$ \vspace{-4pt}\\
   & $7$ & $0.0247495202489046 + 0.0183738440993958i$ \vspace{-4pt}\\
   & $8$ & $0.0210862837711776 + 0.0207122863689407i$ \vspace{-4pt}\\
   & $9$ & $0.0176399327840039 + 0.0222573314442506i$ \vspace{-4pt}\\
   & $10$ & $0.0144010839340230 + 0.0233065697436193i$ \\
\hline
\end{tabular}
\end{subtable}
\hfill
\begin{subtable}{0.49\textwidth}
\caption{$\ell=3$}
\vspace{-5pt}
\begin{tabular}{c@{\hspace{10pt}}c@{\hspace{10pt}}c}
\hline
$\ell$ & $j$ &  $r_H e_j$ \\
\hline
   & $0$ & $0.1006072239125757 + 0.0915010162162569i$ \vspace{-4pt}\\
   & $1$ & $0.0908230933313796 + 0.0251617955322758i$ \vspace{-4pt}\\
   & $2$ & $0.0631457098839271 - 0.0012762849371758i$ \vspace{-4pt}\\
   & $3$ & $0.0405826063112734 - 0.0058654477129917i$ \vspace{-4pt}\\
   & $4$ & $0.0266444759092582 - 0.0027460041524308i$ \vspace{-4pt}\\
$3$& $5$ & $0.0187515720371715 + 0.0014333798543680i$ \vspace{-4pt}\\
   & $6$ & $0.0140972716344782 + 0.0046182669826315i$ \vspace{-4pt}\\
   & $7$ & $0.01097558821497615 + 0.00661974226141833i$ \vspace{-4pt}\\
   & $8$ & $0.00857150329016809 + 0.00773824127797637i$ \vspace{-4pt}\\
   & $9$ & $0.00654988811306071 + 0.00828024569376265i$\vspace{-4pt}\\
   & $10$ & $0.00478503575349286 + 0.00845673795231441i$ \\
\hline
\end{tabular}
\end{subtable}%
\label{tab:n1ell23}
\end{table}


\begin{table}[ht]
\caption{$e_j$ data for $n=2$.}
\vspace{-7pt}
\hspace*{-10pt}
\centering
\begin{subtable}{0.49\textwidth}
\caption{$\ell=2$}
\vspace{-5pt}
\begin{tabular}{c@{\hspace{10pt}}c@{\hspace{10pt}}c}
\hline
$\ell$ & $j$ &  $r_H e_j$ \\
\hline
   & $0$ & $-0.006678041383435 + 0.141460034751257i$ \vspace{-4pt}\\
   & $1$ & $0.1193062609510017 + 0.0756163764534419i$ \vspace{-4pt}\\
   & $2$ & $0.1146657081287942 + 0.0007488463687182i$ \vspace{-4pt}\\
   & $3$ & $0.0782877891066255 - 0.0131338321833861i$ \vspace{-4pt}\\
   & $4$ & $0.0599473129083836 + 0.0012772989022803i$ \vspace{-4pt}\\
$2$& $5$ & $0.0565936649372375 + 0.0160077547135168i$ \vspace{-4pt}\\
   & $6$ & $0.0568488399946155 + 0.0255671324891611i$ \vspace{-4pt}\\
   & $7$ & $0.0560425975806698 + 0.0325413833400877i$ \vspace{-4pt}\\
   & $8$ & $0.0539183634932275 + 0.0390560260774349i$ \vspace{-4pt}\\
   & $9$ & $0.0511353840916677 + 0.0456725141154477i$ \vspace{-4pt}\\
   & $10$ & $0.0480745969771377 + 0.0522805954645665i$ \\
\hline
\end{tabular}
\end{subtable}
\hfill
\begin{subtable}{0.49\textwidth}
\caption{$\ell=3$}
\vspace{-5pt}
\begin{tabular}{c@{\hspace{10pt}}c@{\hspace{10pt}}c}
\hline
$\ell$ & $j$ &  $r_H e_j$ \\
\hline
   & $0$ & $0.0417297228847025 + 0.1054357814502190i$ \vspace{-4pt}\\
   & $1$ & $0.0809446392217919 + 0.0393002156329161i$ \vspace{-4pt}\\
   & $2$ & $0.0657815259440879 - 0.0003182440479682i$ \vspace{-4pt}\\
   & $3$ & $0.0430191259862773 - 0.0089970325320210i$ \vspace{-4pt}\\
   & $4$ & $0.0289927226551848 - 0.0040793839493983i$ \vspace{-4pt}\\
$3$& $5$ & $0.0226844579843885 + 0.0026963752537984i$ \vspace{-4pt}\\
   & $6$ & $0.0197951387879185 + 0.0076629122566710i$ \vspace{-4pt}\\
   & $7$ & $0.0177112285288053 + 0.0108680547155301i$ \vspace{-4pt}\\
   & $8$ & $0.0155449354304499 + 0.0130679060054845i$ \vspace{-4pt}\\
   & $9$ & $0.0131958602719892 + 0.0147600683337322i$ \vspace{-4pt}\\
   & $10$ & $0.0107593575889026 + 0.0161444044960749i$ \\
\hline
\end{tabular}
\end{subtable}%
\label{tab:n2ell23}
\end{table}



\begin{table}[H]
\caption{$e_j$ data for $n = 3$.}
\vspace{-7pt}
\hspace*{-20pt}
\begin{subtable}{0.49\textwidth}
\centering
\caption{$\ell=2$}
\vspace{-5pt}
\begin{tabular}{c@{\hspace{10pt}}c@{\hspace{10pt}}c}
\hline
$\ell$ & $j$ &  $r_H e_j$ \\
\hline
   & $0$ & $-0.0325852551314068 + 0.0747719102635759i$  \vspace{-4pt}\\
   & $1$ & $0.0913442595092874 + 0.0802769838946251i$  \vspace{-4pt}\\
   & $2$ & $0.1221777342069701 + 0.0103635649576794i$  \vspace{-4pt}\\
   & $3$ & $0.0956292086910920 - 0.0116599781393095i$  \vspace{-4pt}\\
   & $4$ & $0.0841321049646752 + 0.0067377139623656i$  \vspace{-4pt}\\
$2$& $5$ & $0.0936680274794272 + 0.0268198345546321i$  \vspace{-4pt}\\
   & $6$ & $0.1063650780631990 + 0.0394793789659170i$  \vspace{-4pt}\\
   & $7$ & $0.1150476659540682 + 0.0513105036772837i$  \vspace{-4pt}\\
   & $8$ & $0.1214885885651855 + 0.0665332576927772i$  \vspace{-4pt}\\
   & $9$ & $0.127970980521597 + 0.084915740662278i$  \vspace{-4pt}\\
   & $10$ & $0.134865872349529 + 0.105123598889745i$ \\
\hline
\end{tabular}
\end{subtable}
\hfill
\begin{subtable}{0.49\textwidth}
\centering
\caption{$\ell=3$}
\vspace{-5pt}
\begin{tabular}{c@{\hspace{10pt}}c@{\hspace{10pt}}c}
\hline
$\ell$ & $j$ &  $r_H e_j$ \\
\hline
   & $0$ & $-0.0015779165948527 + 0.0873367275805243i$  \vspace{-4pt}\\
   & $1$ & $0.0673743381165955 + 0.0486885005868322i$  \vspace{-4pt}\\
   & $2$ & $0.0679844063269548 + 0.0025796969637562i$  \vspace{-4pt}\\
   & $3$ & $0.0461217486198088 - 0.0109948921592254i$  \vspace{-4pt}\\
   & $4$ & $0.0321822975816820 - 0.0046665357787451i$  \vspace{-4pt}\\
$3$& $5$ & $0.0284401186916899 + 0.0044902751481566i$  \vspace{-4pt}\\
   & $6$ & $0.0284145137326224 + 0.0107615638124564i$  \vspace{-4pt}\\
   & $7$ & $0.0281106342904294 + 0.0149900036743692i$  \vspace{-4pt}\\
   & $8$ & $0.0267641215905967 + 0.0187884225796811i$  \vspace{-4pt}\\
   & $9$ & $0.0247640461065528 + 0.0227806369056446i$  \vspace{-4pt}\\
   & $10$ & $0.0224497106473862 + 0.0269194971851201i$ \\
\hline
\end{tabular}
\end{subtable}%
\label{tab:n3ell23}
\end{table}


\begin{table}[H]
\caption{$e_j$ data for $n=4$.}
\vspace{-7pt}
\hspace*{-20pt}
\begin{subtable}{0.49\textwidth}
\centering
\caption{$\ell=2$}
\vspace{-5pt}
\begin{tabular}{c@{\hspace{10pt}}c@{\hspace{10pt}}c}
\hline
$\ell$ & $j$ &  $r_H e_j$ \\
\hline
   & $0$ & $-0.0231849383760028 + 0.0386350792467110i$  \vspace{-4pt}\\
   & $1$ & $0.0745603838858844 + 0.0739685626200772i$  \vspace{-4pt}\\
   & $2$ & $0.1289818292501889 + 0.0185747235596811i$  \vspace{-4pt}\\
   & $3$ & $0.1185869296093861 - 0.0075521097199608i$  \vspace{-4pt}\\
   & $4$ & $0.1208149855689374 + 0.0153836167032158i$  \vspace{-4pt}\\
$2$& $5$ & $0.153246787920723 + 0.042165247302196i$  \vspace{-4pt}\\
   & $6$ & $0.190883591791763 + 0.059087655770938i$  \vspace{-4pt}\\
   & $7$ & $0.223122109713004 + 0.079607772795008i$  \vspace{-4pt}\\
   & $8$ & $0.256242481590652 + 0.111916146131414i$  \vspace{-4pt}\\
   & $9$ & $0.295805911965927 + 0.153900293946673i$  \vspace{-4pt}\\
   & $10$ & $0.341401447538711 + 0.202009507605051i$ \\
\hline
\end{tabular}
\end{subtable}
\hfill
\begin{subtable}{0.49\textwidth}
\centering
\caption{$\ell=3$}
\vspace{-5pt}
\begin{tabular}{c@{\hspace{10pt}}c@{\hspace{10pt}}c}
\hline
$\ell$ & $j$ &  $r_H e_j$ \\
\hline
   & $0$ & $-0.0206699647600537 + 0.0604937732682900i$  \vspace{-4pt}\\
   & $1$ & $0.0531273798618641 + 0.0521831825708380i$  \vspace{-4pt}\\
   & $2$ & $0.0685467879135409 + 0.0065104733363981i$  \vspace{-4pt}\\
   & $3$ & $0.0491845253435823 - 0.0118840345930306i$  \vspace{-4pt}\\
   & $4$ & $0.0357563149149284 - 0.0045009772316074i$  \vspace{-4pt}\\
$3$& $5$ & $0.0357171302438170 + 0.0068894045675336i$  \vspace{-4pt}\\
   & $6$ & $0.0398939005969786 + 0.0139081057618229i$  \vspace{-4pt}\\
   & $7$ & $0.0423766399766888 + 0.0189679808169785i$  \vspace{-4pt}\\
   & $8$ & $0.0429977895228710 + 0.0251497406761657i$  \vspace{-4pt}\\
   & $9$ & $0.0430899787206310 + 0.0330274154954972i$  \vspace{-4pt}\\
   & $10$ & $0.0431505747879606 + 0.0419485811820150i$ \\
\hline
\end{tabular}
\end{subtable}%
\label{tab:n4ell23}
\end{table}


\begin{table}[H]
\caption{$e_j$ data for $n=5$.}
\vspace{-9pt}
\hspace*{-20pt}
\begin{subtable}{0.49\textwidth}
\centering
\caption{$\ell=2$}
\vspace{-5pt}
\begin{tabular}{c@{\hspace{10pt}}c@{\hspace{10pt}}c}
\hline
$\ell$ & $j$ &  $r_H e_j$ \\
\hline
   & $0$ & $-0.0094305225379713 + 0.0228491067858833i$  \vspace{-4pt}\\
   & $1$ & $0.0702008814396801 + 0.0662882049778363i$  \vspace{-4pt}\\
   & $2$ & $0.142252676129377 + 0.025405358939820i$  \vspace{-4pt}\\
   & $3$ & $0.154620812604551 - 0.000423219329531i$  \vspace{-4pt}\\
   & $4$ & $0.182203230561820 + 0.029504645326924i$  \vspace{-4pt}\\
$2$& $5$ & $0.255336456159544 + 0.066808671315109i$  \vspace{-4pt}\\
   & $6$ & $0.342726500675614 + 0.092694584974768i$  \vspace{-4pt}\\
   & $7$ & $0.430371143569840 + 0.130949071521098i$  \vspace{-4pt}\\
   & $8$ & $0.533689309697611 + 0.196920669397476i$  \vspace{-4pt}\\
   & $9$ & $0.665552920988299 + 0.285553231284859i$  \vspace{-4pt}\\
   & $10$ & $0.824806651834691 + 0.390298807833414i$ \\
\hline
\end{tabular}
\end{subtable}
\hfill
\begin{subtable}{0.49\textwidth}
\centering
\caption{$\ell=3$}
\vspace{-5pt}
\begin{tabular}{c@{\hspace{10pt}}c@{\hspace{10pt}}c}
\hline
$\ell$ & $j$ &  $r_H e_j$ \\
\hline
   & $0$ & $-0.0242359941938532 + 0.0391998561347797i$ \vspace{-4pt}\\
   & $1$ & $0.0410525578207275 + 0.0510883587925916i$  \vspace{-4pt}\\
   & $2$ & $0.0674611376093846 + 0.0101191863850904i$  \vspace{-4pt}\\
   & $3$ & $0.0518041022306143 - 0.0121462820686405i$  \vspace{-4pt}\\
   & $4$ & $0.0394451464476780 - 0.0038860226006234i$  \vspace{-4pt}\\
$3$& $5$ & $0.0443110947175641 + 0.0097199793849433i$  \vspace{-4pt}\\
   & $6$ & $0.0542554531049167 + 0.0169330906459240i$  \vspace{-4pt}\\
   & $7$ & $0.0608334954196972 + 0.0226710726902531i$  \vspace{-4pt}\\
   & $8$ & $0.0652918023688233 + 0.0323530592228630i$  \vspace{-4pt}\\
   & $9$ & $0.0706834627858642 + 0.0461483209546148i$  \vspace{-4pt}\\
   & $10$ & $0.0774416624771747 + 0.0622664396956380i$ \\
\hline
\end{tabular}
\end{subtable}%
\label{tab:n5ell23}
\end{table}

\begin{table}[H]
\caption{$e_j$ data for $n=6$.}
\vspace{-8pt}
\hspace*{-10pt}
\begin{subtable}{0.49\textwidth}
\centering
\caption{$\ell=2$}
\vspace{-5pt}
\begin{tabular}{c@{\hspace{10pt}}c@{\hspace{10pt}}c}
\hline
$\ell$ & $j$ &  $r_H e_j$ \\
\hline
   & $0$ & $0.0071189800022886 + 0.0244270829307704i$ \vspace{-4pt}\\
   & $1$ & $0.0791909802110347 + 0.0620357577486420i$ \vspace{-4pt}\\
   & $2$ & $0.173115614615052 + 0.035388200761421i$ \vspace{-4pt}\\
   & $3$ & $0.222901483433564 + 0.015778312404325i$ \vspace{-4pt}\\
   & $4$ & $0.301561328505558 + 0.059893181498572i$ \vspace{-4pt}\\
$2$& $5$ & $0.455611505703941 + 0.118981345967461i$ \vspace{-4pt}\\
   & $6$ & $0.651812379158736 + 0.169875613740407i$ \vspace{-4pt}\\
   & $7$ & $0.876783760510318 + 0.251723052069271i$ \vspace{-4pt}\\
   & $8$ & $1.165440235050287 + 0.392822370971920i$ \vspace{-4pt}\\
   & $9$ & $1.54894032266647 + 0.58575879231756i$ \vspace{-4pt}\\
   & $10$ & $2.03128586446286 + 0.82238818017992i$ \\
\hline
\end{tabular}
\end{subtable}
\hfill
\begin{subtable}{0.49\textwidth}
\centering
\caption{$\ell=3$}
\vspace{-5pt}
\begin{tabular}{c@{\hspace{10pt}}c@{\hspace{10pt}}c}
\hline
$\ell$ & $j$ &  $r_H e_j$ \\
\hline
   & $0$ & $-0.0218337427685166 + 0.0255734369366163i$ \vspace{-4pt}\\
   & $1$ & $0.0320875572902621 + 0.0477002188031165i$ \vspace{-4pt}\\
   & $2$ & $0.0655086920252467 + 0.0127396297373017i$ \vspace{-4pt}\\
   & $3$ & $0.0540292854986211 - 0.0122365023143462i$ \vspace{-4pt}\\
   & $4$ & $0.0432800997782630 - 0.0031200844577528i$ \vspace{-4pt}\\
$3$& $5$ & $0.0542418031712442 + 0.0127726879121193i$ \vspace{-4pt}\\
   & $6$ & $0.0717092006682787 + 0.0196507330512797i$ \vspace{-4pt}\\
   & $7$ & $0.0840206147652311 + 0.0259611517154740i$ \vspace{-4pt}\\
   & $8$ & $0.0950486968143402 + 0.0406293154981293i$ \vspace{-4pt}\\
   & $9$ & $0.1108277658771737 + 0.0628258921132427i$ \vspace{-4pt}\\
   & $10$ & $0.131373960775933 + 0.088831277287416i$ \\
\hline
\end{tabular}
\end{subtable}%
\label{tab:n6ell23}
\end{table}

\begin{table}[H]
\caption{$e_j$ data for $n=7$.}
\vspace{-8pt}
\hspace*{-10pt}
\begin{subtable}{0.49\textwidth}
\centering
\caption{$\ell=2$}
\vspace{-5pt}
\begin{tabular}{c@{\hspace{10pt}}c@{\hspace{10pt}}c}
\hline
$\ell$ & $j$ &  $r_H e_j$ \\
\hline
   & $0$ & $0.0287995454850320 + 0.0184008711381451i$ \vspace{-4pt}\\
   & $1$ & $0.1185260892634616 + 0.0702388413679524i$ \vspace{-4pt}\\
   & $2$ & $0.262798164747494 + 0.067651801946992i$ \vspace{-4pt}\\
   & $3$ & $0.402418479961720 + 0.075446166908563i$ \vspace{-4pt}\\
   & $4$ & $0.619265957172561 + 0.167123015914277i$ \vspace{-4pt}\\
$2$& $5$ & $0.990929923676543 + 0.300580618898559i$ \vspace{-4pt}\\
   & $6$ & $1.50132244234922 + 0.45516305413662i$ \vspace{-4pt}\\
   & $7$ & $2.16022900341441 + 0.69861773253752i$ \vspace{-4pt}\\
   & $8$ & $3.05660702313977 + 1.08904787182772i$ \vspace{-4pt}\\
   & $9$ & $4.28089213354471 + 1.62947163418331i$ \vspace{-4pt}\\
   & $10$ & $5.88191594348812 + 2.32700575812425i$ \\
\hline
\end{tabular}
\end{subtable}
\hfill
\begin{subtable}{0.49\textwidth}
\centering
\caption{$\ell=3$}
\vspace{-5pt}
\begin{tabular}{c@{\hspace{10pt}}c@{\hspace{10pt}}c}
\hline
$\ell$ & $j$ &  $r_H e_j$ \\
\hline
   & $0$ & $-0.0181695599769233 + 0.0173653805837275i$ \vspace{-4pt}\\
   & $1$ & $0.0258006211401533 + 0.0436621925138659i$ \vspace{-4pt}\\
   & $2$ & $0.0633760140904404 + 0.0144010604651561i$ \vspace{-4pt}\\
   & $3$ & $0.0560671363333610 - 0.0123560243951517i$ \vspace{-4pt}\\
   & $4$ & $0.0474063165779652 - 0.0023395700887393i$ \vspace{-4pt}\\
$3$& $5$ & $0.0656394145006476 + 0.0159386609575311i$ \vspace{-4pt}\\
   & $6$ & $0.0925735912492434 + 0.0219612321451617i$ \vspace{-4pt}\\
   & $7$ & $0.1126404955138170 + 0.0287748178212681i$ \vspace{-4pt}\\
   & $8$ & $0.133997197586964 + 0.050300439609534i$ \vspace{-4pt}\\
   & $9$ & $0.167594131545889 + 0.083835192044083i$ \vspace{-4pt}\\
   & $10$ & $0.212604721378107 + 0.122574246944150i$ \\
\hline
\end{tabular}
\end{subtable}%
\label{tab:n7ell23}
\end{table}

\begin{table}[H]
\caption{$e_j$ data for $n=8$.}
\vspace{-8pt}
\begin{subtable}{0.49\textwidth}
\centering
\caption{$\ell=2$}
\vspace{-5pt}
\begin{tabular}{c@{\hspace{10pt}}c@{\hspace{10pt}}c}
\hline
$\ell$ & $j$ &  $r_H e_j$ \\
\hline
   & $0$ & ${1164765 i}/{700009}$ \vspace{-4pt}\\
   & $1$ & ${5571603 i}/{1400018}$ \vspace{-4pt}\\
   & $2$ & ${5823825 i}/{700009}$ \vspace{-4pt}\\
   & $3$ & ${44416947 i}/{2800036}$ \vspace{-4pt}\\
   & $4$ & ${19775313 i}/{700009}$ \vspace{-4pt}\\
$2$& $5$ & ${66780435 i}/{1400018}$ \vspace{-4pt}\\
   & $6$ & ${3176547 i}/{41177}$ \vspace{-4pt}\\
   & $7$ & ${674140005 i}/{5600072}$ \vspace{-4pt}\\
   & $8$ & ${127575708 i}/{700009}$ \vspace{-4pt}\\
   & $9$ & ${188176863 i}/{700009}$ \vspace{-4pt}\\
   & $10$ & ${271337670 i}/{700009}$ \\
\hline
\end{tabular}
\end{subtable}
\hfill
\begin{subtable}{0.49\textwidth}
\centering
\caption{$\ell=3$}
\vspace{-5pt}
\begin{tabular}{c@{\hspace{10pt}}c@{\hspace{10pt}}c}
\hline
$\ell$ & $j$ &  $r_H e_j$ \\
\hline
   & $0$ & $-0.0147928320971783 + 0.0123743181062329i$ \vspace{-4pt}\\
   & $1$ & $0.0214391560286977 + 0.0397521519745856i$ \vspace{-4pt}\\
   & $2$ & $0.0614090924597511 + 0.0153637244331753i$ \vspace{-4pt}\\
   & $3$ & $0.0580838091121277 - 0.0125425795554257i$ \vspace{-4pt}\\
   & $4$ & $0.0519502219622454 - 0.0015730406072024i$ \vspace{-4pt}\\
$3$& $5$ & $0.0786457141341650 + 0.0191862568338425i$ \vspace{-4pt}\\
   & $6$ & $0.1172032879552634 + 0.0238326931912907i$ \vspace{-4pt}\\
   & $7$ & $0.147509636765554 + 0.031110164267112i$ \vspace{-4pt}\\
   & $8$ & $0.184171013412476 + 0.061763756771027i$ \vspace{-4pt}\\
   & $9$ & $0.245875062224800 + 0.110035571897353i$ \vspace{-4pt}\\
   & $10$ & $0.330574124897447 + 0.164396535331579i$ \\
\hline
\end{tabular}
\end{subtable}%
\label{tab:n8ell23}
\end{table}

\begin{table}[H]
\caption{$e_j$ data for $n=9$.}
\vspace{-7pt}
\hspace*{-20pt}
\begin{subtable}{0.49\textwidth}
\centering
\caption{$\ell=2$}
\vspace{-5pt}
\begin{tabular}{c@{\hspace{10pt}}c@{\hspace{10pt}}c}
\hline
$\ell$ & $j$ &  $r_H e_j$ \\
\hline
   & $0$ & $-0.0481759684369441 + 0.0198318150040291i$ \vspace{-4pt}\\
   & $1$ & $-0.0857100274407277 + 0.0696760632695613i$ \vspace{-4pt}\\
   & $2$ & $-0.163048616819628 + 0.088879078933767i$ \vspace{-4pt}\\
   & $3$ & $-0.399175255262569 + 0.12223674006529i$ \vspace{-4pt}\\
   & $4$ & $-0.81065760615737 + 0.260087843129221i$ \vspace{-4pt}\\
$2$& $5$ & $-1.42480435859151 + 0.46296090697546i$ \vspace{-4pt}\\
   & $6$ & $-2.45012481988937 + 0.70044231531924i$\vspace{-4pt}\\
   & $7$ & $-4.13067788001983 + 1.09884440280146i$ \vspace{-4pt}\\
   & $8$ & $-6.62226314194639 + 1.77828164952022i$ \vspace{-4pt}\\
   & $9$ & $-10.16840189854991 + 2.71688284859888i$ \vspace{-4pt}\\
   & $10$ & $-15.263695475771 + 3.8936799351803i$ \\
\hline
\end{tabular}
\end{subtable}
\hfill
\begin{subtable}{0.49\textwidth}
\centering
\caption{$\ell=3$}
\vspace{-5pt}
\begin{tabular}{c@{\hspace{10pt}}c@{\hspace{10pt}}c}
\hline
$\ell$ & $j$ &  $r_H e_j$ \\
\hline
   & $0$ & $-0.0120283376057713 + 0.0092268293347515i$ \vspace{-4pt}\\
   & $1$ & $0.0183816535851371 + 0.0362400948763753i$ \vspace{-4pt}\\
   & $2$ & $0.0597273795213759 + 0.0158699469639609i$ \vspace{-4pt}\\
   & $3$ & $0.0601776636210653 - 0.0127784268842468i$ \vspace{-4pt}\\
   & $4$ & $0.0570000117323253 - 0.0008112575506905i$ \vspace{-4pt}\\
$3$& $5$ & $0.0933947167540892 + 0.0225146379534043i$ \vspace{-4pt}\\
   & $6$ & $0.145978216615856 + 0.025264285239297i$ \vspace{-4pt}\\
   & $7$ & $0.189560074188723 + 0.032997533643411i$ \vspace{-4pt}\\
   & $8$ & $0.247936809811029 + 0.075464851825739i$ \vspace{-4pt}\\
   & $9$ & $0.351473842723456 + 0.142352794591404i$ \vspace{-4pt}\\
   & $10$ & $0.496761868293284 + 0.215162791465753i$ \\
\hline
\end{tabular}
\end{subtable}%
\label{tab:n9ell23}
\end{table}

\begin{table}[H]
\caption{$e_j$ data for $n=10$.}
\vspace{-7pt}
\hspace*{-20pt}
\begin{subtable}{0.49\textwidth}
\centering
\caption{$\ell=2$}
\vspace{-5pt}
\begin{tabular}{c@{\hspace{10pt}}c@{\hspace{10pt}}c}
\hline
$\ell$ & $j$ &  $r_H e_j$ \\
\hline
   & $0$ & $-0.0286426058703113 + 0.0106113087976747i$ \vspace{-4pt}\\
   & $1$ & $-0.0441254333176308 + 0.0464183237687842i$ \vspace{-4pt}\\
   & $2$ & $-0.0752048505169943 + 0.0468937958675964i$ \vspace{-4pt}\\
   & $3$ & $-0.227596261935816 + 0.042132374231758i$ \vspace{-4pt}\\
   & $4$ & $-0.508076137978851 + 0.118357621671413i$ \vspace{-4pt}\\
$2$& $5$ & $-0.915575048516598 + 0.224172274528923i$ \vspace{-4pt}\\
   & $6$ & $-1.64633503018129 + 0.29889860678943i$ \vspace{-4pt}\\
   & $7$ & $-2.94384311234699 + 0.46714791896406i$ \vspace{-4pt}\\
   & $8$ & $-4.91739505425679 + 0.85380608133462i$ \vspace{-4pt}\\
   & $9$ & $-7.75386071925519 + 1.37454445656736i$\vspace{-4pt}\\
   & $10$ & $-11.97130226833821 + 1.91254938844106i$ \\
\hline
\end{tabular}
\end{subtable}
\hfill
\begin{subtable}{0.49\textwidth}
\centering
\caption{$\ell=3$}
\vspace{-5pt}
\begin{tabular}{c@{\hspace{10pt}}c@{\hspace{10pt}}c}
\hline
$\ell$ & $j$ &  $r_H e_j$ \\
\hline
   & $0$ & $-0.00983772607665567 + 0.00715383021451585i$ \vspace{-4pt}\\
   & $1$ & $0.0162007321968787 + 0.0331738592666763i$ \vspace{-4pt}\\
   & $2$ & $0.058347734545312 + 0.016090541198903i$ \vspace{-4pt}\\
   & $3$ & $0.0624036576883143 - 0.0130377492445573i$ \vspace{-4pt}\\
   & $4$ & $0.0626211364925363 - 0.0000384220075813i$ \vspace{-4pt}\\
$3$& $5$ & $0.1100219464703823 + 0.0259314159606847i$ \vspace{-4pt}\\
   & $6$ & $0.17931606216785 + 0.026268228832864i$ \vspace{-4pt}\\
   & $7$ & $0.239862602404814 + 0.034485599094529i$ \vspace{-4pt}\\
   & $8$ & $0.328047625741178 + 0.091885300951663i$ \vspace{-4pt}\\
   & $9$ & $0.491232700613689 + 0.181775433171764i$ \vspace{-4pt}\\
   & $10$ & $0.725014956209716 + 0.275712379374871i$ \\
\hline
\end{tabular}
\end{subtable}%
\label{tab:n10ell23}
\end{table}

\end{document}